\documentclass[12pt, preprint]{aastex}

\slugcomment{To appear AJ August 27 2002}

\shorttitle{The extremely metal poor CH star \SarasStar}
\shortauthors{Lucatello {\it et al.}}
\newcommand{\kms}{km~s$^{-1}$}
\newcommand{\SarasStar}{HE~0024$-$2523}
\newcommand{\CompStar}{HE~2344$-$2800}
\newcommand{\teff}{$T_{\mbox{\scriptsize eff}}$}

\begin{document}

\title{Stellar Archaeology: a Keck Pilot Program on Extremely Metal-
Poor  Stars From the Hamburg/ESO Survey. III. 
The Lead (Pb) Star \SarasStar \altaffilmark{1, 2}}

\author{Sara Lucatello\altaffilmark{3, 4},  
Raffaele Gratton\altaffilmark{3}, Judith G. Cohen\altaffilmark{5},
 Timothy C. Beers\altaffilmark{6},  
Norbert Christlieb\altaffilmark{7,8,9}, Eugenio Carretta\altaffilmark{3}
 \& Solange Ram\'{\i}rez \altaffilmark{5}}

\altaffiltext{1}{Based in part on observations obtained at the
        W.M. Keck Observatory,  which is operated jointly by the California
        Institute of Technology,  the University of California
        and NASA.}
\altaffiltext{2}{Based in part on observations collected at the European
Southern Observatory,  Paranal,  Chile (ESO Programme 167.D-0173(A)).}
\altaffiltext{3}{INAF, Osservatorio Astronomico di Padova,  Vicolo dell'Osservatorio 5, 
        35122,  Padova,  Italy.}
\altaffiltext{4}{Dipartimento di Astronomia,  Universit\'a di Padova,  Vicolo dell'Osservatorio 2, 
        35122,  Padova,  Italy.}
\altaffiltext{5}{Palomar Observatory,  Mail Stop 105-24, 
        California Institute of Technology,  Pasadena,  CA \,  91125.}
\altaffiltext{6}{Department of Physics and Astronomy,  Michigan State University, 
East Lansing,  Michigan 48824-1116.}
\altaffiltext{7}{Hamburger Sternwarte,  Gojenbergsweg 112,  D-21029,  Hamburg, 
Germany.}
\altaffiltext{8}{Department
   of Astronomy and Space Physics, Uppsala University, Box 524,
   SE-75239 Uppsala, Sweden.}
\altaffiltext{9}{Marie Curie Fellow, on sabbatical leave from Hamburger
   Sternwarte.}

\begin{abstract}
We present a detailed abundance analysis, including spectral syntheses, of a
very metal-poor ([Fe/H] $\simeq -2.7$), peculiar main sequence star\,(\SarasStar)
detected during the course of the Keck Pilot Program.  Radial velocities of
this star were obtained during four different observing runs over a time span
of 1.1 years, and demonstrate that it is clearly a short period spectroscopic
binary. An orbital solution was obtained, and orbital parameters were
determined with high precision. The rotational velocity was also measured
($v_{rot}\sin i=9.7\pm1.5$\,\kms); rotation appears likely to be synchronous
with the orbit.

The abundance analysis and spectral syntheses indicate that the object is a CH
star characterized by extreme {\it s}-process enrichment, likely
due to mass accretion from an evolved companion which has now probably
become a white dwarf. The lead (Pb) abundance of \,(\SarasStar) is very high, 
the same as that of the recently discovered lead-rich metal-poor star
CS~29526-110, [Pb/Fe]$=+3.3$.
The abundance ratio of the heavy-{\it s} to light-{\it s} elements, as
characterized by Pb and Ba, [Pb/Ba]$=+1.9$, is the highest yet found for any
metal-poor star, and is about 0.7\,dex higher than that of CS~29526-110.  On
the basis of the measured isotopic ratio of carbon (C$^{12}$/C$^{13}\sim$ 6) we
argue that the mass donor must have had an original mass of at least $\sim
3$\,${\rm M}_{\odot}$. The unusually short period of this CH star suggests that
it underwent a past common-envelope phase with its evolved companion.  

Our results are compared to the latest available models for AGB yields and {\it
s}-process nucleosynthesis.  We also discuss the possible connection between
\SarasStar, the lithium depletion of halo stars, and halo blue straggler
formation.

\end{abstract}
\section{Introduction \label{intro}}

One of the surprising results of the hitherto largest wide-field spectroscopic
survey for metal-poor stars, the HK survey (Beers {\it et al.} 1992; Beers
1999), is the high frequency of carbon-enhanced stars among very metal-poor
stars.  Norris {\it et al.} (1997a) found that 14\,\% of the HK survey stars
with $\mbox{[Fe/H]}<-2.5$\footnote{We use the usual spectroscopic notation:
$\log$ n(A) is the abundance (by number) of the element A on the usual scale
where $\log$ n(H)=12, while [A/H] denotes the logarithmic ratio of the
abundances of elements A and H in the star, minus the same quantity in the
Sun.} have stronger than normal G bands. More recent
results, based on a larger sample, suggest that the frequency of
carbon-enhanced stars ({\it i.e.}, stars with $\mbox{[C/Fe]}> +1.0$) among
metal-poor stars with $\mbox{[Fe/H]}<-2.5$ is perhaps as high as 25\,\%.
These findings are confirmed by the Hamburg/ESO objective-prism survey, HES herafter,
(Wisotzki {\it et al.} 2000; Christlieb {\it et al.} 2001a,b), where $\sim
30\,\%$ of the metal-poor candidates appear to be  carbon-enhanced stars.

Despite extensive investigations of this class of objects by means of
high-resolution spectroscopy (Norris {\it et al.} 1997a,b; Bonifacio {\it et
al.} 1998; Hill {\it et al.} 2000; Aoki {\it et al.} 2002a; Norris {\it et
al.} 2002), the origin of carbon in these stars still remains unclear. The
extremely metal-poor star CS~22892$-52$ (McWilliam {\it et al.} 1995,
$\mbox{[Fe/H]}=-3.1$) exhibits an overabundance of carbon
($\mbox{[C/Fe]}=+1.0$) as well as of r-process elements (Sneden {\it et al.}
1994; 1996; 2000;  $\mbox{[r/Fe]}\sim +1.6$), hinting at a common mechanism
for the enrichment, in particular, by supernovae of Type II (SNII), as
discussed by Aoki {\it et al.} (2002a). However, Aoki et al. also found a class
of carbon-rich, metal-poor stars without neutron capture element excess,
arguing against a common enrichment mechanism. Furthermore, a second,
recently-discovered very metal-poor star with strong r-process enrichment,
CS~31082$-$001 (Cayrel {\it et al.} 2001), is only very slightly
carbon-enhanced, $\mbox{[C/Fe]}=+0.2$ (Hill {\it et al.} 2002).


A different scenario for the origin of carbon in these stars is that proposed
for the moderately metal-poor, classical CH stars ($\mbox{[Fe/H]}\sim
-1.5$), in which the carbon-enhanced star is a member of a wide binary
system that accreted material from a former primary, during the asymptotic
giant branch (AGB) phase of the latter, as described by McClure \& Woodsworth
(1990). However, while all CH stars are indeed members of binary systems
(McClure \& Woodsworth 1990), and radial velocity variations have been
reported (and periods determined) for some carbon-enhanced metal-poor
(hereafter CEMP) stars (Norris {\it et al.} 1997a, Aoki {\it et al.} 2000,
Preston \& Sneden 2001), the latter authors found that three of their CEMP
subgiants do not show any radial velocity variations greater than
0.4\,km\,s$^{-1}$ over a period of 8 years, strongly arguing against the mass-transfer
scenario for these three stars. Furthermore, in this scenario it is
expected that the star we observe today should display an enrichment of {\it
s}-process elements, produced by the former primary in its AGB phase, while the
CEMP star CS~22957$-$027 (Norris {\it et al.} 1997b; Bonifacio {\it et al.}
1998), as well as the previously mentioned class of stars found by Aoki et al.,
do not exhibit this behavior.

The CEMP stars that \emph{do} show {\it s}-process enrichment provide the
unique opportunity to study an extinct generation of extremely metal-poor AGB
stars that left their fingerprints on their less massive companions. These
stars provide strong observational constraints for theoretical models of the
structure, evolution, and nucleosynthesis of early-epoch AGB stars, and permit
studies of the {\it s}-process operating at very low metallicities.  Recent
calculations by Goriely \& Siess (2001) predict that efficient production of
{\it s}-process elements takes place even at zero metallicity, despite the
absence of iron seeds, provided that protons are mixed into carbon-rich layers.
Proton mixing results in the formation of $^{13}$C, which is a strong neutron
source due to the reaction $^{13}$C($\alpha$, n)$^{16}$O. The recent discovery
of CEMP stars with strong overabundances of Pb by Aoki {\it et al.} (2000) and
Van Eck {\it et al.} (2001) are in concert with these predictions.

In two previous papers, Cohen {\it et al.} (2002, hereafter Paper I)  and
Carretta {\it et al.} (2002, hereafter Paper II), we analyzed a sample of
extremely metal-poor stars from the Hamburg/ESO prism survey. As
briefly discussed in Paper~II, one of the objects, \SarasStar, exhibited a
chemical signature typical of CH stars. The G band of CH was found to be
extremely strong, while the evidence of a large overabundance of
neutron-capture elements, coupled with an apparent lack of strong Eu lines,
suggested that the object was enriched by the {\it s}-process. Furthermore, the
previously reported detection of the Pb line at $4057.8$\,{\AA}  suggested an
enormous lead overabundance. In order to measure the abundances of carbon, the
{\it s}-process elements, and the light element N with higher precision, and to
study the dynamical behavior of this peculiar object, more observations were
needed.  In this paper we present results from an analysis of several
spectra obtained with the Keck~I, VLT-UT2, and the Palomar Hale telescopes.

After summarizing our observations in \S \ref{obsdr}, we describe the radial
velocity measurements, the modeling of the radial velocity curve, and the
computation of the orbital parameters in \S \ref{dyn}. In \S \ref{checomp} we
present the abundance analysis from equivalent widths and 
from spectral syntheses.  We discuss our 
interpretation of these results in \S \ref{disc}.
\section{Observations and data reductions \label{obsdr}}

A total of 18 spectra of \SarasStar\, have been collected during several runs
using three different instrument-telescope combinations: UVES at VLT-UT2 Kueyen
(Dekker {\it et al.} 1997), HIRES at Keck I (Vogt {\it et al.} 1994), and the
Double Spectrograph at the Hale 5-m Telescope at Palomar Mountain (Oke \& Gunn,
1982).

The UVES observations were carried out during two different runs, on August
27$^{th}$ 2001, and between October 23$^{rd}$ and 27$^{th}$ 2001. Three
metal-poor field stars, G4-37, G126-62 and G141-19, were observed with the same
setup, so that they could be used as templates to measure rotational
velocities.  The signal-to-noise ratio for the template spectra is $\sim 200/1$
per resolution element. The spectra have a resolution of
$R=\lambda/\Delta\lambda\simeq50,000$, where $\Delta\lambda$ refers to the
width of the resolution elements.  A dichroic beam-splitter was employed, which
distributed the light gathered with the telescope to the blue and red arms of
UVES, covering a range from 3600 to 4800\,{\AA} and from 5700 to 9500\,\AA,
respectively. The slit width was fixed at 1 arcsec and the 
binning
 used 1$\times$2, where 1 refers to the dispersion direction and
2 to the spatial one.  The spectra were extracted and wavelength calibrated using the
standard UVES pipeline. The bias and scattered light were subtracted from the
spectra, flat-fielding was applied, and the extracted spectra were wavelength
calibrated with arc-lamp spectra taken during the same observing
run; then the orders were blaze corrected and merged.
 
The Keck I HIRES spectra are described in detail in Paper I. They have a wavelength
coverage from 3850 to 5400\,\AA, and were observed with a slit width of 0.86
arcsec and 1$\times$2 binning on the CCD chip. The first five spectra were all
collected during a single night, September 24$^{th}$ 2000. Two more spectra
were obtained, one on the 27$^{th}$, and the other on the 30$^{th}$ of
September 2001. All the spectra have a resolution of $R\simeq 45,000$, and
their extraction and wavelength calibration were performed using FIGARO
(Shortridge, 1993). The reduction was similar to that performed on the UVES
spectra, but in this case the orders were not merged.

The Palomar spectra were obtained during four consecutive nights, from October
24$^{th}$ to October 27$^{th}$ 2001, with a 1.0 arcsec slit and 0.6 \AA\,pixels,
yielding a spectral resolution of $R\simeq $3,500 over the spectral region 6100
to 6700 \AA, with H$\alpha$ the major spectral feature present in that range.
FIGARO was again used to extract and wavelength calibrate the spectra. Several
radial velocity standards (HD~2796, HD~38230, HD~171232, and CS~22950-046) were
also observed in order to test the reliability of our measurements. Further
details on the observations are listed in Table \ref{obs}.
%
%
%

\section{Dynamics \label{dyn}}

\subsection{Data analysis \label{datad}}

\subsubsection{Radial velocities \label{vrd}}

Radial velocities were measured by fitting Gaussians to several prominent lines
in different orders  and adding the appropriate barycentric correction. The
values measured for the radial velocity standards showed a systematic offset in
the Palomar spectra of $\sim$4.8\,\kms, which  was added to the measured radial
velocities as a correction for instrumental effects. The measured heliocentric
radial velocities are given in Table \ref{obs}.

\subsubsection{Rotational velocities \label{vrot}}

As mentioned in Paper I, the spectra of \SarasStar\, were characterized by
broader lines than those of the other program stars, suggesting a non-
negligible rotational velocity. In Paper I, we roughly estimated
v$_{\mbox{\scriptsize rot}}$  to be $\sim$ 8\,\kms.

Here we improve upon this measurement. The rotational velocity was measured
using a cross-correlation technique; the spectrum of the program star was
cross-correlated against a few template spectra of non-rotating stars with
similar atmospheric parameters. The cross-correlation peaks were fit with a
Gaussian curve; from the resulting FWHM the value of $v \sin i$ was calculated
according to a relation calibrated on several other stars with well-known
rotational velocities.  A more detailed description of the procedure used can
be found in Lucatello \& Gratton (2002). The spectrum used for this purpose was
that with the highest signal-to-noise ratio, the UVES spectrum taken on
August 27$^{th}$ 2001. Only the blue part was used to measure rotation, as it
is not affected by telluric lines and exhibited the strongest signatures in the
cross-correlations. The spectrum was first divided into single orders.
Each order was then cross-correlated against the corresponding orders
of the three different templates using the IRAF task {\it rv.fxcor}.  From the
resulting average FWHMs, after subtracting all the {\it external} sources of
broadening ({\it e.g.} template contribution, instrumental broadening, etc.),
we obtained a value of {\it v $\sin i$} = 9.7$\pm$1.5\,\kms\ . This value is
strictly an upper limit, as it includes also non-rotational intrinsic stellar
broadening, such as macro-turbulence.  However, as shown in Lucatello \&
Gratton (2002), this procedure provides a good estimate of the value of the
true {\it v $\sin i$} when its value is larger than $\sim 6$~\kms.

\subsection{Modeling the radial velocity curve \label{rvc}}

In order to compute the orbital parameters for \SarasStar , which for the sake
of clarity we will refer to as the {\it primary star} hereafter (the companion
will be called the {\it secondary}),  we employed the program {\it Orbitsolver}
by J. S.  Varga, G. A. Radford, D. W. Beggs, R. E. M. Griffin and J. R.
Lewis (R. E. M. Griffin, priv. comm.). This program adjusts all the elements
of the orbit simultaneously in a least-squares solution (Griffin \& Cornell,
1998).

The measurements of the radial velocities were fed into the program, along with
the corresponding MJD of each of the observations. The values of the velocities
were weighted according to their errors of observation.  However, note that
we did not use the individual errors, but rather an estimate of the average
errors, $\sim 1$\,\kms\, for HIRES and UVES data, and $\sim 5$\,\kms\, for the
Palomar data.  The initial values of the parameters were provided from
direct examination of the data.  The initial period was assumed to be four
days, since we noticed that the radial velocity was very similar for the two
UVES observations taken at MJD= 52206.014 and  52210.021.  For the initial
semi-amplitude of the orbit we used half the difference between the highest and
lowest velocity points, and for the epoch at zero phase we assumed the epoch of
the first observation. A stable and reasonable solution was obtained after five
iterations.

\begin{figure}
\plotone{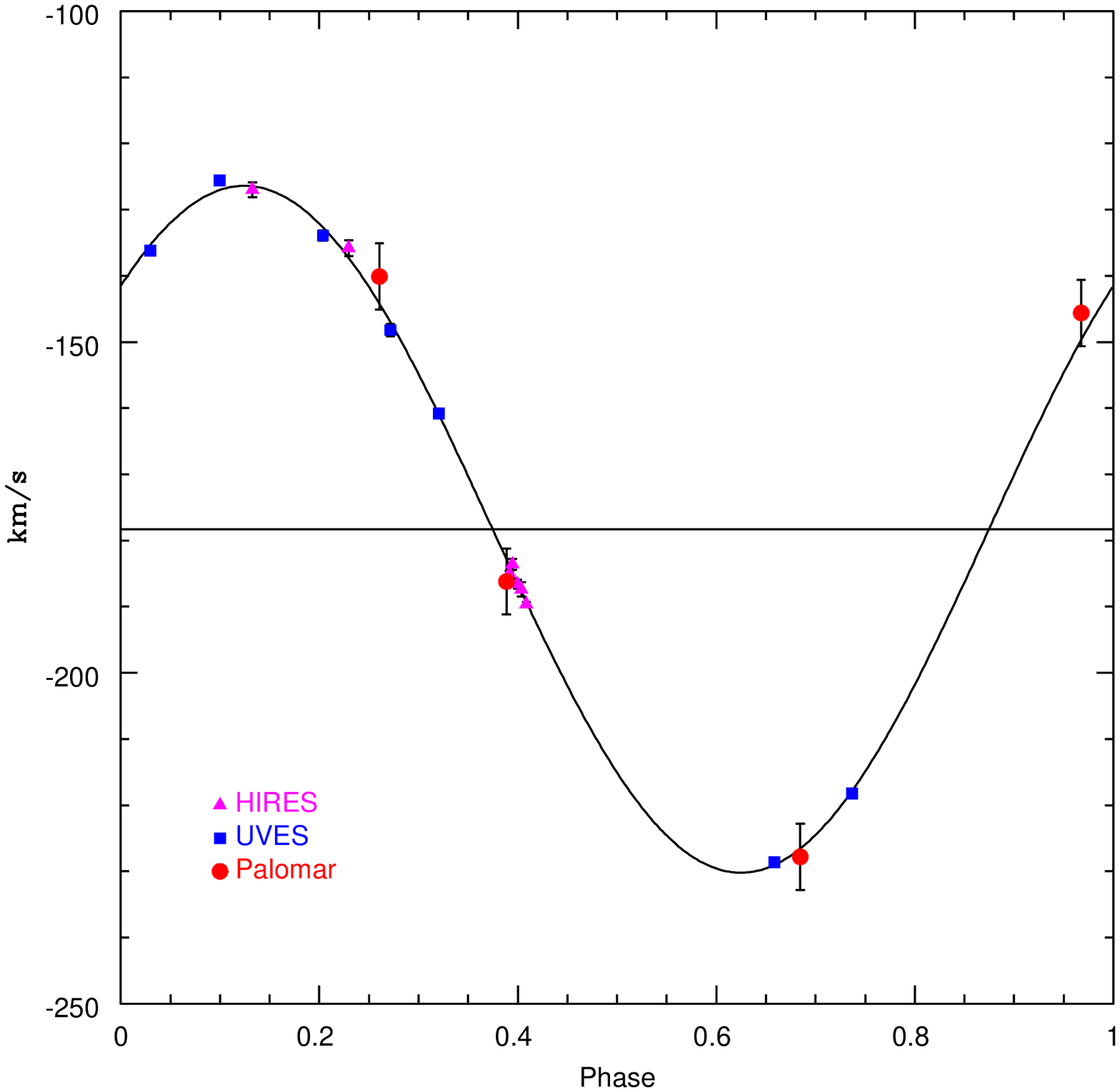}
\caption{Radial velocity curve. The solid curve represents our {\it best}
solution, the symbols are the data phased according to the period we derive
(see Table \ref{orbsol} for details of the orbital parameters for our orbital
solution). Closed triangles are HIRES data and closed squares are UVES data
while Palomar data are shown as filled circles. The radial velocity errors are
shown for the Palomar spectra, where the lower spectral dispersion leads to
much larger uncertainties. The HIRES/UVES spectra have errors
comparable to the size of the symbols used. The solid line indicates  the
radial velocity of the barycenter. \label{radvelcur}}
\end{figure}

Initially, we assumed a circular orbit ({\it i.e.} {\it e}=0); after having
refined the other orbital elements, we allowed the eccentricity to vary as
well. Our best solutions for circular and non-circular orbits are provided
in Table \ref{orbsol}, along with their values before the last iteration. The
short period suggests that the orbit should have an eccentricity close to zero.
%
%
%
%
Our computation of the orbital elements, solving for {\it e} as well, obtained
a very small value for {\it e}, and did not significantly change the other
orbital parameters with respect to those obtained when adopting a circular
orbit.  Allowing the eccentricity to vary did not improve the solution
significantly.  In fact, the circular solution yielded  smaller 
r.m.s. residuals, hence we adopted a circular orbit.  Figure
\ref{radvelcur} shows the radial-velocity data together with the adopted
solution, while Table \ref{vr} lists the observed values for the radial
velocities along with their expected value from our orbital solution.
%
%
%

\subsection{Mass assumptions and consequences on the orbital parameters
\label{massass}}

Although the values of the individual masses are not known, we can derive
useful limits on them through considerations based on the available
photometric and spectroscopic information for \SarasStar .

\subsubsection{{\it Primary} star mass \label{primarymass}} 

We begin by considering the mass of the primary.  If we assume a 
constant spatial density for the extreme halo population, which is reasonable
for distances up to a few \,kpc, and a uniform mass function (adopted since
information on the mass of the observed star is lacking)\footnote{Note that
these assumptions are not critical in the present discussion}, then the
probability of observing a star of a given mass at the measured magnitude and
color can be expressed by the formula:
\begin{equation}
P(M^{*})=\frac{\sum_{i=1}^2 \tau_{i}(M^{*}) \Delta V_{i}(M^{*})}{\sum_{j=1}^{n} \sum_{i=1}^2
\tau_{i}(M_{j}) \Delta V_{i}(M_{j})}
\end{equation}
where $\tau_{i}(M_{j})$ is the time spent in the {\it i}th phase
(with i=1 being the main-sequence phase and i=2 the subgiant phase) by the star
of {\it j}-th mass within the restricted evolutionary phase allowed by the
atmospheric parameters. $\Delta${\it V(M)} is the volume sampled by stars of
the measured luminosity, which on turn depends on the intrinsic luminosity of
the object.  The integral in the denominator normalizes $P(M)$.

In order to compute these quantities we used  the Y$^{2}$ tracks (Yi {\it et al.}
2001) for a mass interval 0.4$<$ M/M$_{\odot}<$5 with 0.1\,M$_{\odot}$ bins, with
a metallicity [Fe/H]$=-3.3$, and [$\alpha$/Fe]$=+0.3$.  We assumed the
$\pm$2\,$\sigma$ range around the measured temperature of \SarasStar\,({\it
{\it i.e.}} 6425 to 6825\,K) to specify the evolutionary phase, and computed the
time spent in this range. In this way we obtained two permitted intervals for
each mass,  one on the main sequence and the other along the subgiant branch.
The number of stars of the corresponding mass observable at the measured
magnitude was calculated from the theoretical tracks, assuming a bolometric
magnitude for the Sun of $M_b = 4.711$ and a bolometric correction for our star
of BC(V)$=-0.091$ (Lejeune {\it et al.} 1998).

\begin{figure}
\centering
\plotone{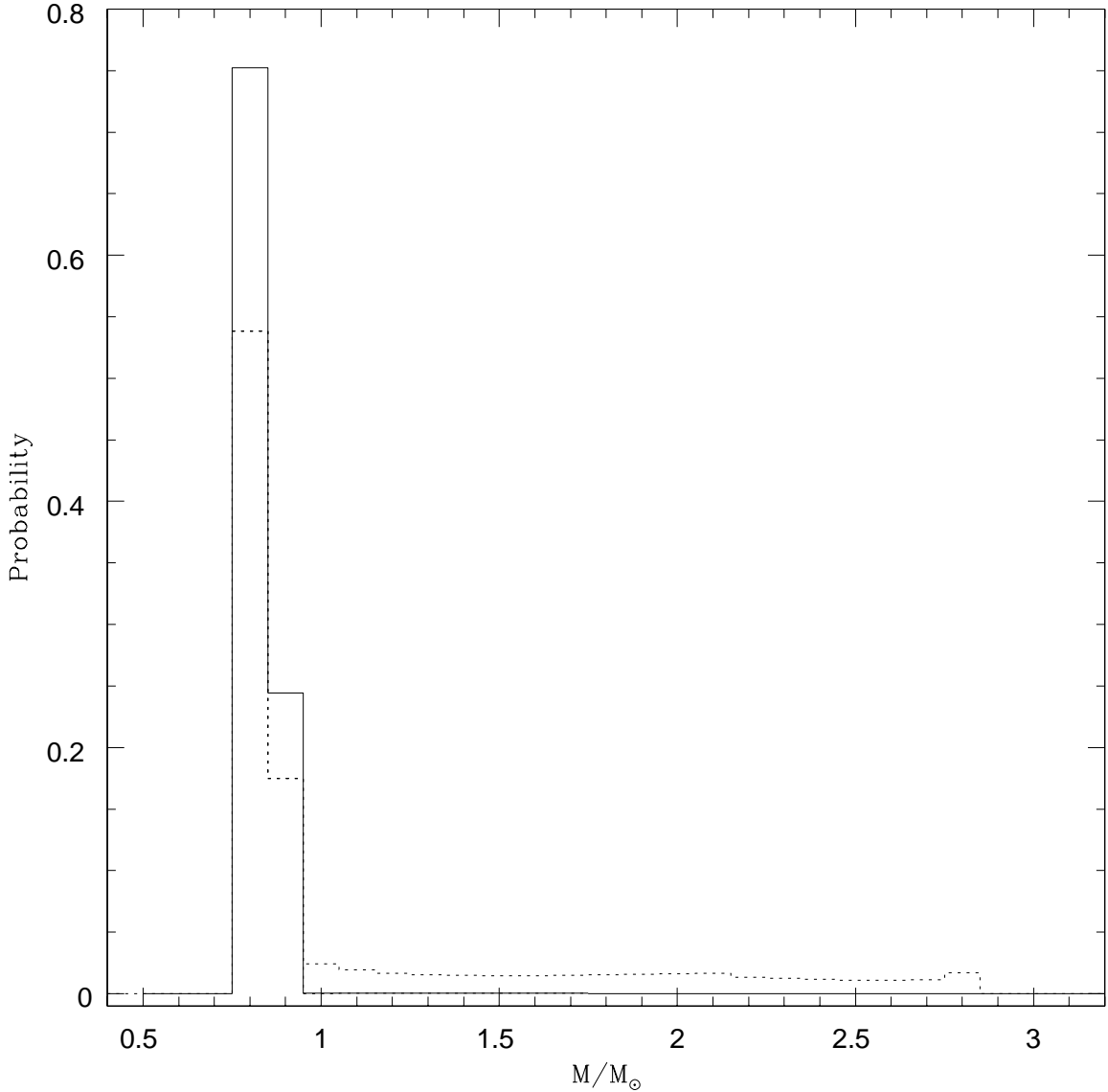}
\caption{
The solid line histogram represents the normalized probability distribution
for the mass of the observed star. It has been computed using the Y$^{2}$ tracks
(Yi {\it et al.} 2001) and assuming constant stellar density (for a detailed
description see text). The tracks for M$>$2.8\,M$_{\odot}$ do not cross the
allowed region of \teff\, and hence have a probability of 0.  The
probability function drops abruptly at 0.9\,M$_{\odot}$, but seems to be
approximately flat in the interval  0.9$<$M/M$_{\odot}<$2.8. This is due to
the fact that as the time spent in the observed temperature region decreases,
the intrinsic luminosity increases, which increases the volume of space in
which stars of such luminosity are observed at the measured magnitude, and so
the product of the two terms is roughly constant. The dashed line histogram
represents the probability distribution for the mass range allowed by the
requirements on the gravity. \label{prob}}
\end{figure}

The final probabilities were evaluated according to Eq. 1; Figure \ref{prob}
provides a graphical representation of the results. In this figure a value of
zero for the probability of a given mass indicates that the track for that mass
does not cross the allowed region in \teff. This figure shows that the
probability function drops abruptly at 0.9\,M$_{\odot}$, but seems to be
roughly constant (at low but non-zero values) in the interval $0.95 < {\rm
M}/{\rm M}_{\odot} < $2.85.  This behavior arises because as the time spent in
the observed temperature region decreases, the intrinsic luminosity increases,
hence the volume of space in which stars of such luminosity are
observed at the measured magnitude increases.
%
%
%
Application of the ionization equilibrium for Fe supplies a further constraint,
since it limits the acceptable values of surface gravity.  The lower limit 
on the range of gravity may be obtained from the difference
between the derived abundances of Fe\,II and Fe\,I.  This difference is (see
Table \ref{abu}) [Fe\,II/H]-[Fe\,I/H]$=0.05\pm0.12$, where the uncertainty
is the quadratic sum of $\sigma$ of the measured Fe\,II abundance (that of
Fe\,I, being 0.02\,dex, is negligible when added in quadrature), and of the
error on the Fe\,I abundance arising from uncertainties in the temperature
(see Table 5 of Paper II and the discussion in \S 4.2). Thus, the maximum
difference between the abundances of the ionized and neutral iron species is,
at a 2\,$\sigma$ confidence level, ([Fe/H]II-[Fe/H]I)$_{max}=0.28$.
According to the sensitivity of the abundances to the atmospheric parameters
given in Table 5 of Paper II, this corresponds to a lower limit for the gravity
of $\log (g)=3.7$\,dex.

Taking into account this constraint, the probability for masses larger than
0.95\,M$_{\odot}$ is reduced to about 0.003 (since only the low-luminosity
phase is acceptable for larger masses). Thus, on the basis of these simple
statistical considerations, it is  reasonable to assume that the mass of the
primary lies in the range  0.75 $<$M/M$_{\odot}< 0.95$.

\subsubsection{{\it Secondary} star mass \label{secondarymass}}

It is very likely that the {\it secondary} star has undergone thermal pulses
during its AGB phase, as we observe the expected nucleosynthesis
products (see \S \ref{abuan} and following): the {\it s}-process elements that
we detect in the {\it primary} star must have been produced in the {\it
secondary},   as the {\it primary} seems to be presently on the main sequence.
The AGB-nucleosynthesis-enriched material was then transfered to the {\it
primary} star,   polluting its atmosphere. The minimum amount of material
required to create the signatures characteristic of CH stars is quite small,
as the polluted star is very metal poor.

The mass of the {\it secondary}, which is now likely a white dwarf, is likely
to be consistent with the mass of the degenerate CO core of the star in its AGB
phase. Thus, based on this assumption, it is possible to obtain a reliable
lower limit to the mass of this star.  The minimum core mass for a star to
experience third dredge-up is a (theoretically) very well-studied quantity
over the metallicity range $-1.3<$[Fe/H]$<0$ (Herwig {\it et al.} 1997;
Boothroyd \& Sackman 1988a,b,c,d ; Marigo, Girardi \& Bressan 1999). However,
no detailed computation has been performed for metallicities as low as that of
\SarasStar, hence extrapolation to such low metallicities might be premature.

Since we seek an upper limit for the value of $\sin i$, high precision
is not crucial; a reasonable estimate of this quantity can provide a
meaningful constraint.  Hence we adopt the value of 0.6\,M$_{\odot}$ from an
{\it ad-hoc} computation (Straniero 2002,  priv. comm.).

\subsubsection{Orbital inclination and synchronous rotation \label{incl}}

Inversion of the expression for the mass function,
$f(m)=\frac{m_{2}^{3}\sin^{3}i}{(m_{1}+m_{2})^{2}}$\ (which is known from the
orbital solution), and substitution, for $m_{1}$ and $m_{2}$, respectively, of
the upper and lower limits estimated in the previous sections, we
obtain:
\begin{equation}
\sin i\leq\sqrt[3]{\frac{(m_{1}+m_{2})^{2}}{m_{2}^{3}} f(m)}=0.82
\end{equation}
which corresponds to an inclination of {\it i} $\leq55.1^{\circ}$. It should be
kept in mind that this is an upper limit to the angle and {\it not} an
estimate of the angle itself. It is noteworthy that the value of  $\sin i$ is
far more sensitive to the mass of the {\it secondary} than it is to the
{\it primary}.

On the basis of this upper limit on the orbital inclination angle, we can
derive an upper limit to the value of the synchronous rotational velocity. A
star is in a synchronous orbit if its rotational period is the same as its
period of revolution.  Since the orbital period has been measured, we obtain,
using the derived upper limit on $\sin i$ for the upper limit of the projected
value of the rotational velocity:
\begin{equation}
v{\mbox{\scriptsize
rot,syn}}\sin i\leq\frac{2 \pi {\rm R} \sin i}{{\rm P}}
\end{equation}
where $v{\mbox{\scriptsize
rot,syn}}$ is the upper limit of the value of the synchronous rotational
velocity, R is the radius of the star, and P is the orbital revolution period.
Using the same Y$^{2}$ (Yi {\it et al.}, 2001) isochrone as that described in
Paper I, the value for the radius corresponding to the atmospheric parameters
adopted was computed to be ${\rm R}=1.10\,{\rm R}_{\odot}$. Substituting this
value in Eq. 3, we obtain an upper limit for the synchronous projected
rotational velocity of $~v{\mbox{\scriptsize rot,syn}}\sin
i\leq13.5\pm3.9$\,\kms. The uncertainty is mainly due to the errors in the
determination of the radius, since the period and the mass function are
determined with a much higher degree of accuracy. The error of the radius
depends on the uncertainties of the effective temperature and gravity. Such
uncertainties are, respectively, of about $\pm$100 K in \teff\, (which
translates into an uncertainty of $10\%$ on the radius), and, as discussed in
\S 3.3.1, of $\pm$0.3\,dex in $\log (g)$ (about $25\%$ in the radius).
The total error on the radius, and consequently on the $v{\mbox{\scriptsize
rot,syn}}\sin i$, is $29\%$ (1 $\sigma$) .

The measured value of $v_{rot} \sin i=9.7 \pm 1.5$\,\kms, is compatible with
the computed upper limit. Since the value of the orbital semi-major axis is
only a few times larger than the inferred radius, it is likely that tidal
forces led the system to reach a state of synchronicity and a circular orbit,
the latter of which has been derived from our observations.  Thus the orbit
can be considered synchronous and the observed $v_{rot} \sin i$ the true
synchronous rotational velocity of the {\it primary}.

Assuming synchronicity, we can derive the value of $\sin i$, hence a
relationship between the values of the masses of the two components. This
value is subject to the error computed for the upper limit of the angle (see
Eq. 2), plus that due to that of the {\it measured} rotational velocity, which
is $\pm 15\%$. In this manner, using Eq. 3 and substituting $v_{rot} \sin i$
for $v{\mbox{\scriptsize rot,syn}}\sin i$, we obtain $\sin i$=0.59$\pm$0.19,
corresponding to an angle of {\it i}=($36\pm^{15}_{7}$)$^{\circ}$.
 
The expression of the mass function links the values of the {\it primary} and
the {\it secondary} masses, so that once one of them is known, under the
hypothesis of a synchronous orbit, the other is uniquely determined. However,
in this case this computation does not add any information, as the mass range of
the {\it secondary} corresponding to that computed for the {\it primary} in
\S 3.3.1 -- calculated from the mass function using for 
$\sin i$ the maximum, minimum and mean value
allowed in the case of synchronous orbit  --  covers the entire interval from 0.6\,M$_{\odot}$ to
 (and beyond) the
Chandrasekar limit (see Table \ref{mass}). Thus, the only limits we are
able to derive are $0.75\,M_{\odot}<m_{1}<0.95\,M_{\odot}$\ and
$0.60\,M_{\odot}<m_{2}<1.4\,M_{\odot}$.
%
%
%
\section{Chemical composition \label{checomp}}

Paper II described the results for the chemical composition of \SarasStar\,
from the HIRES spectra taken in September 2000. In the following sections we
present the results obtained from the abundance analysis from lines equivalent widths 
 (\S \ref{abuan})
and spectral syntheses (\S \ref{synt}) for both the UVES and HIRES spectra. For
both purposes we used the UVES spectrum with the highest signal-to-noise ratio
({\it i.e.} that taken on  August 27$^{th}$ 2001) and the sum of the HIRES
spectra obtained in September 2000. The addition of the UVES spectrum to our
data sample improves the signal-to-noise ratio, and permits the detection and
measurement of lines that were not within the spectral range covered by the
HIRES spectra.

\subsection{Atmospheric parameters \label{paratmo}} 

The procedure used to determine the atmospheric parameters for \SarasStar\,
have been thoroughly explained in Paper I: \teff\, is derived from broad-band
colors,  taking the mean of the de-reddened colors $V-K$ and $V-J$; once
\teff\, is specified, $\log$($g$) is obtained from the Y$^{2}$ isochrones.

With \teff\ and $\log$($g$) fixed, [Fe/H] is determined interactively by
matching the observed equivalent widths (EWs) to those computed by
integrating the line profiles of the adopted atmospheric model,  obtained from
the equation of transport computed at different wavelengths along each line.
The model atmosphere we adopt was extracted by interpolation from the Kurucz
grid of models (Kurucz CD-ROM 18, 1995) without overshooting. This model assumes
a scaled solar composition: while this is not strictly appropriate in the case
of \SarasStar, modifications in the model structure are not very large. A crude
estimate of these modifications on the model atmosphere due to the high carbon
abundance ([C/Fe]=+2.6) can be obtained with the following method. We
measured the integrated absorption due to all of the spectral lines in the
range between 4300 and 4335\,\AA~, both in the program star and in
another star of the Keck Pilot Program (\CompStar), which has almost identical
atmospheric parameters. This interval was chosen as it is one of the spectral
ranges most affected by the presence of CH. The resulting ratio of the
summed absorption integrated over this spectral range 
is  $\sim 2.5$. As a very conservative estimate\footnote{Even
after having taken into account the contribution due to other CH bands, CN
bands and CO bands, this estimate is essentially an upper limit. This spectral
range is one of the most affected by a high carbon abundance since it covers
most of the G-band of CH absorption. A more realistic value would be $\sim$
2.}, we assumed that the CH absorption affects the model in the same way a
spectral interval which accounts for {\it ten} times ($\sim 0.05$ of the flux,
obtained from the integration of the equation of the flux of a black body at
the same temperature as \SarasStar) as much flux as that emitted in the range
4300 and 4335\,\AA~($\sim 0.005$ of the flux). On this basis, {\it i.e.} that
the remaining spectral regions accounting for about 95\% of the flux are
negligibly affected, we deduced that the difference in the flux of the star
with [C/H]=+2.6, with respect to that with scaled solar carbon abundance, is
$\sim 7.5\%$. At a metallicity of [Fe/H]=$-$2.7, this translates into an
error of 0.03\,dex, which is negligible compared to the errors of the
abundances due to the uncertainties in the atmospheric parameters 
themselves\footnote{Using a different test, Aoki {\it et al.} (2002a) found that the
enhancements in C and N have only a small effect on the model atmospheres in
the temperature range 4000-6000\,K. For an actual calculation of
the effects on C and N excesses on the atmospheric structure we refer the
reader  Hill {\it et al.} (2000).}.

\begin{figure}
\plotone{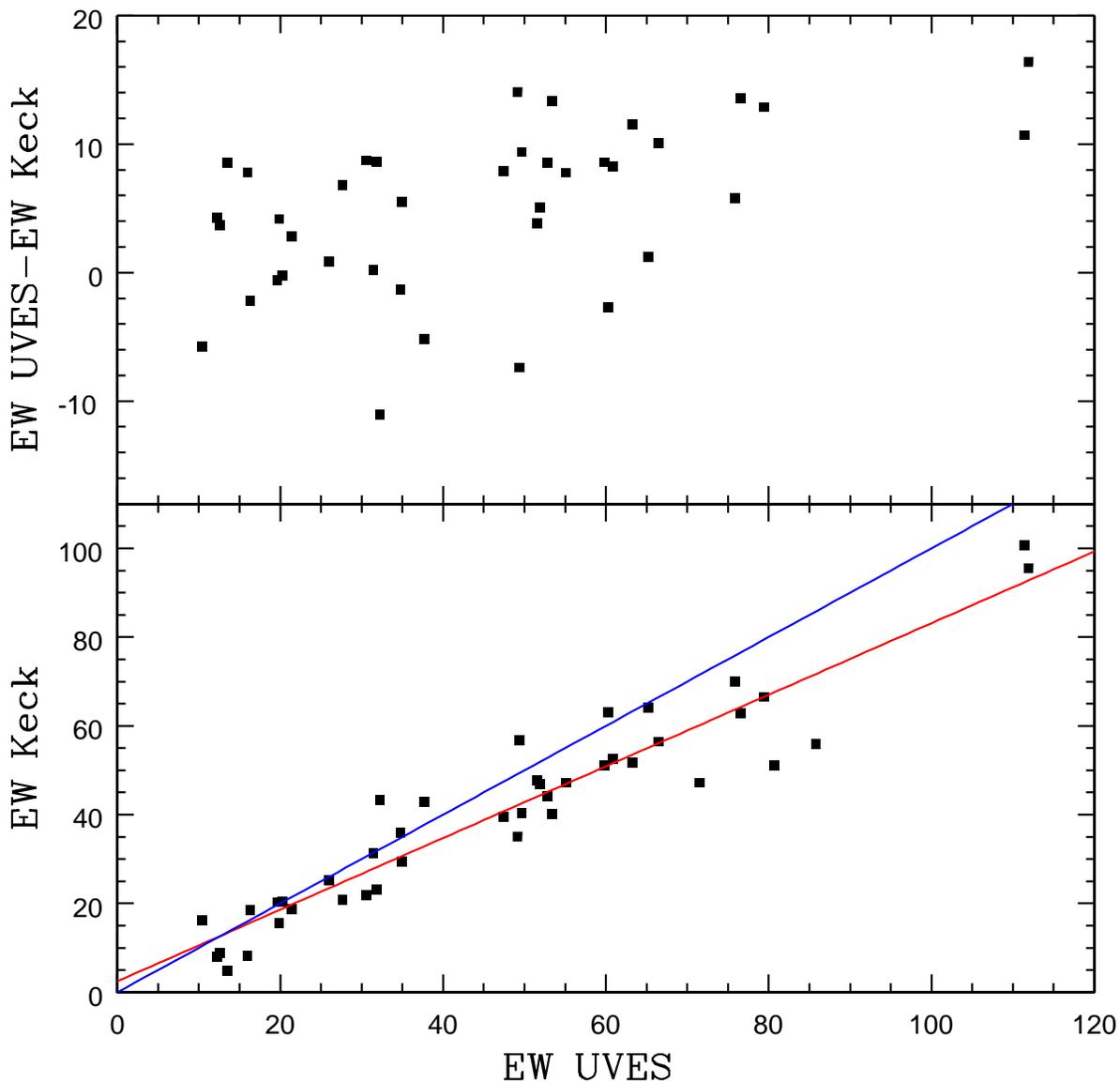}
\caption{Comparison between the EWs measured from the UVES spectrum with those
measured from the HIRES September 2000 spectra. In the bottom panel we plotted the
UVES EWs vs those of HIRES; the upper line is described by the equation 
 EW$_{HIRES}$=EW$_{UVES}$, while the lower line is the best fit and its equation of
the line is EW$_{HIRES}$=(0.82$\pm$0.04){\rm EW}$_{UVES}$+(1.56$\pm$2.10); the
r.m.s. of the fit is 6.8 m\AA. The top panel shows the difference between the
corresponding EWs vs the UVES EWs.\label{ew}}
\end{figure}

\begin{figure}
\plotone{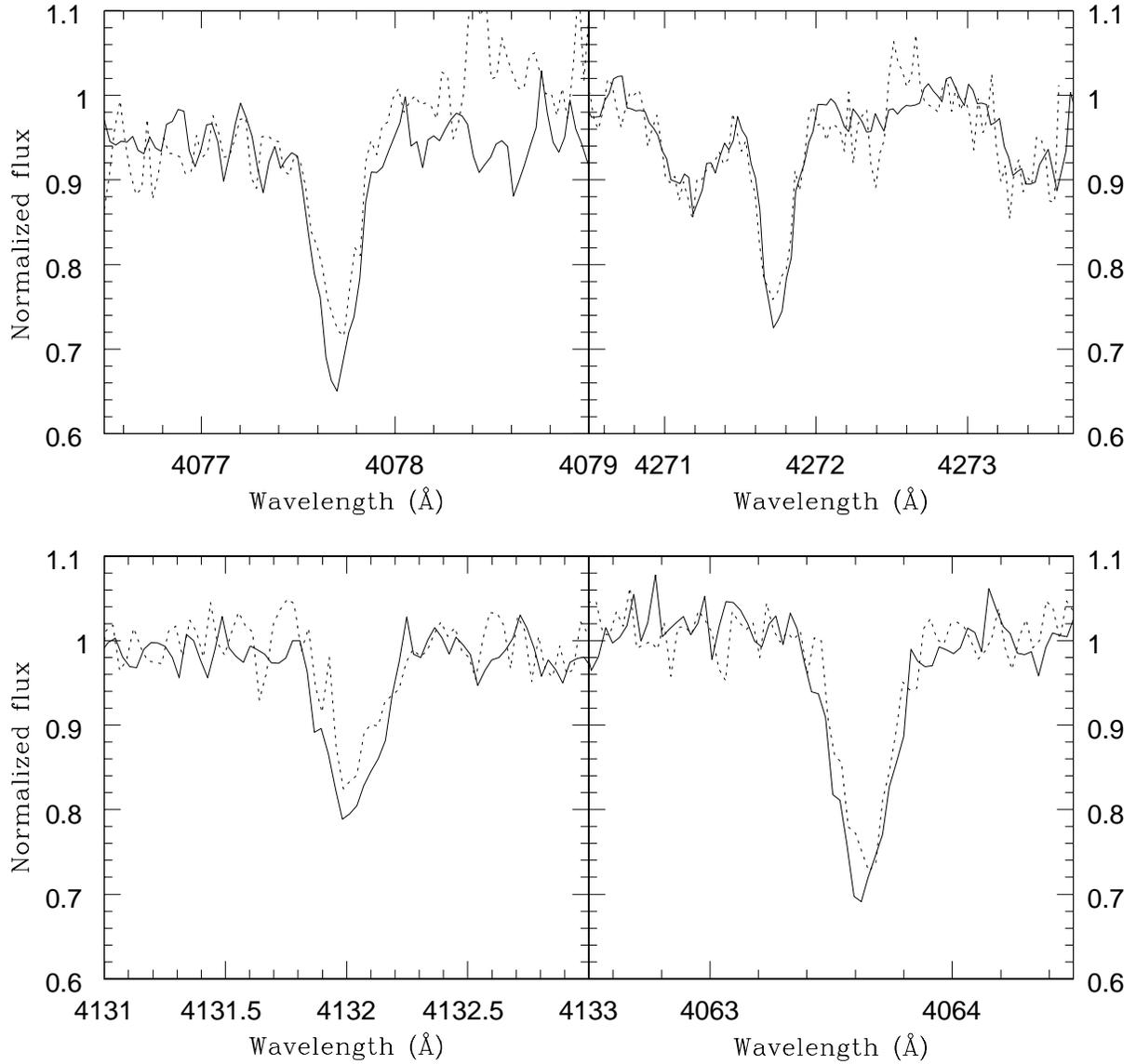}
\caption{ Comparison between four spectral regions from the UVES spectrum
(solid line) and HIRES September 2000 spectra (dashed line). Note that the lines
appear slightly stronger in the UVES spectra. \label{lines}}
\end{figure}

Equivalent widths for the UVES spectrum were measured in the same way as
described in Paper II for the HIRES spectra. Since the resolution of the UVES
spectrum is almost identical to that of the HIRES one,  the signal-to-noise
ratio only slightly better, and the procedure used for the measurements and
selection of EWs are identical, we expect the internal errors to be similar to
those evaluated in Paper II.  The same follows for the external errors (see \S
3 of Paper II). A comparison between the EWs obtained from the UVES spectrum
with those from the HIRES spectra showed that the EWs from the VLT spectra are
systematically larger than those measured from the Keck spectra,  as can be
seen in the top panel of Figure \ref{ew}. The lower panel shows the UVES EWs
{\it vs.} those obtained from HIRES; the upper line is described by the equation ${\rm
EW}_{{\rm HIRES}}={\rm EW}_{{\rm UVES}}$,  while the lower line is the best-fit
relation, described by the equation:
\begin{equation}
{\rm EW}_{{\rm HIRES}}=(0.82 \pm 0.04){\rm EW}_{{\rm UVES}}+(1.56 \pm 2.10)\,{\rm m\AA}.
\end{equation}
%
%
%
%
The r.m.s. scatter around the regression line is 6.8 m\AA; assuming that both
sets have equal errors, we can estimate a typical error of 4.8 m\AA\,for our
measurements in a single spectrum. This difference in the EWs does not depend
on the measurement procedure or on the continuum tracing. To show this, we
compared several strong lines by superimposing some sample regions from the two
spectra in Figure \ref{lines}; the difference between the two spectra is
directly visible there.
%
%
%
%

A contribution due to inadequate subtraction of scattered light during the
reduction does not seem likely. In the UVES spectra, the O$_{2}$ telluric lines
around 7630\,\AA\,reach down to zero intensity, indicating a proper treatment
of scattered light. Furthermore, the comparison between the EWs for
HD 140283 from an HIRES spectrum and those from the literature (see Paper II)
showed no systematic difference, suggesting that, at least in this case,
scattered light was subtracted accurately. We have no explanation at present
for this difference, although it is noteworthy that the spectra were taken at
different phases (spaced roughly by a quarter of an orbit).

Since not all the lines were detected and measured in both spectra, in order
to calculate the EWs to be used for the analysis,  we computed an ``average''
correction,   {\it i.e.} two equations that allow the calculation of the
expected mean value of the EWs from the value measured in only one of the
spectra. The coefficients of the two equations were calculated on the basis of
the lines detected in both spectra, and assuming that the expected value of the
mean EW depends linearly on EWs measured in only one of the spectra. We
obtain:
\begin{equation}
	{\rm EW}_{{\rm mean}}=(1.05 \pm 0.03){\rm EW}_{{\rm HIRES}} + (1.41 \pm 1.21)\,{\rm m\AA} 
\end{equation}
\begin{equation}
	{\rm EW}_{{\rm mean}}=(0.91 \pm 0.02){\rm EW}_{{\rm UVES}} + (0.78 \pm 1.06)\,{\rm m\AA}.
\end{equation} 	
Thus, in the abundance analysis we used the mean of the EWs when a line was
detected and measured in both spectra, or a value corrected using Eq. 5
or 6 if the EW was measured, respectively, only in the HIRES or the UVES
spectrum.

The micro-turbulent velocity was derived by elimination of any trend in
abundances derived from Fe I lines with the expected EW. The final value
adopted herein turned out to be different from that of Paper I,  as more lines
from UVES spectra were added.  In the present analysis, we find \teff=6625\,K
${\rm [Fe/H]}$=$-$2.72, log(g)=4.3\,dex and $v_{\mbox{\scriptsize
micro}}$=1.4\,\kms, compared to ${\rm[Fe/H]}$=$-$2.65\ and
$v_{\mbox{\scriptsize micro}}$=0.58\,\kms\ in Paper II.

The list of EWs, along with the excitation potential and assumed $\log (gf)$ of
the lines is provided in Table \ref{ewt}.
%
%
%

\subsection{Abundance analysis \label{abuan}}

The abundance analysis was performed using the values of the EWs described in
the previous section and listed in Table \ref{ewt}. The resulting abundance and
element ratios for each species are listed in Table \ref{abu},  along with the
number of lines used in the abundance analysis from equivalent width
measurements of a given ion and  $\sigma$, the r.m.s scatter in abundance
obtained from individual lines within the set of lines used for that particular
ion. The abundances of neutral species are computed with respect to Fe I, while
singly ionized species are compared to Fe II, to minimize the effect due to the
choice of atmospheric parameters.
%
%
%

In Paper II we evaluated the sensitivity of the derived abundances to the
atmospheric parameters for the method used, and the results are displayed in
Table 6 of that paper. Since the uncertainties on the atmospheric parameters
and the method used for the abundance analysis are the same, these results
still hold for \SarasStar . The cumulative errors, which take into account the
scatter in the abundances derived from the lines and the effect of the
uncertainties on the atmospheric parameters, are given in Column 5 of Table
\ref{abu}.

\subsubsection{Abundances \label{abures}}

The solar abundances we have assumed to compute the abundance 
ratios are the same as in Paper II, {\it i.e.} those obtained from the 
analysis of the solar spectrum using Kurucz's models (CD-ROM 18, 1995).
These results are very similar to those of Anders \& Grevesse \cite{and}, with the exception 
of iron, which is lower by about 0.1\,dex than their photospheric abundance value, 
while it roughly agrees with their meteoritic value.
The results obtained from the abundance
analysis (see Table \ref{abu}) are quite similar to those presented in Paper
II. The detection of the Zr~II line at 4149.2\,{\AA} and of the Y~II line at
4883.7\,\AA\,are noteworthy additions. Since the signal-to-noise ratio in those
regions was not very high, and we did  not detect any other lines of these
elements, even if other lines of similar strength are expected to be
detectable, we considered those measured EWs as upper limits. The difference
between the measured abundances of Ti~I and Ti~II might suggest that the
assumed \teff\, is too high; however the Fe~I and Fe~II abundances are very
similar and seem to invalidate  this hypothesis. On the other hand, the three
Ti~I lines detected are very weak and the abundances derived from them should
be attributed low weight.

Paper II provides a thorough and detailed description of the sources of
uncertainties in the derived abundance for Fe, describes the oscillator
strengths assumed, and a
comparison with a purely spectroscopic analysis.  A similar discussion is also
provided for the $\alpha$-elements; we refer the interested reader to sections
4.1 and those that follow in the above mentioned paper for a complete
discussion of these aspects of the analysis. 
A basic assumption in our procedure is that departures from LTE are
negligible for Fe. While there is no doubt that Fe II lines form very
close to LTE, Th{\' e}venin \& Idiart (1999) proposed that
there is a significant over-ionization of Fe in metal-poor stars.
According to their predictions, an LTE analysis should underestimate
the abundances derived from Fe I lines by about 0.3 dex in stars with
atmospheric parameters similar to \SarasStar. While such an error would
have no major impact on the conclusions of this paper, we remark that there
is no evidence for such a large error in the Fe ionization equilibrium
from our data (this difference would corresponds to an error of
about 1.2 dex in the surface gravity, much larger than that compatible with the
location of the star which is close to the turn-off). On the oter side, the
prediction by Th{\' e}venin \& Idiart are not confirmed by other statistical
equilibrium computations. Gratton {\it et al.} (1999) estimated a
much smaller non-LTE correction of about 0.1 dex for stars with atmospheric
parameters similar to \SarasStar, but they also noticed that this value
has to be considered as an upper limit to real non-LTE corrections. The
result by Gratton et al. is based on a comparison of expected departures
from LTE with empirical determinations of Fe ionization equilibrium for
stars where departures from LTE are expected to be much stronger than in
turn-off stars, due to the combination of a stronger ionizing flux and low
surface gravity, in particular RR Lyrae. More recently, analysis of even
hotter stars on the blue horizontal branch of several globular clusters (at
temperatures of about 8,000\,K) by Behr (1999) also confirms
that a good equilibrium of ionization for Fe can be obtained assuming LTE
(and atmospheric parameters derived from the location of the star in the
colour-magnitude diagram). According to Gratton et al., non-LTE corrections
for such stars are expected to be a factor of at least 5 larger than in
turn-off stars. We conclude that corrections for departures from LTE should
be small for Fe, and we will neglect them hereinafter.

The Al abundance has been corrected to account for NLTE  effects
using values interpolated from Table 1 of Baum\"uller \& Gehren \cite{bau}.
Hyperfine structure (HFS) was taken into account for Mn (data from Booth, 
Shallis,  \& Wells 1983) and Ba, for which we used data from Steffen
\cite{ste}.

\subsection{Spectral synthesis \label{synt}} 

Abundances for Li, C, N, Eu, La, and Pb were derived from spectral synthesis,
because features due to these elements are either very weak or somewhat
blended with nearby lines due to other species. As a general rule, line lists
used in these spectral syntheses were derived as follows. We extracted from
Kurucz's database (Kurucz CD-ROM 23, 1995) lines of neutral and singly ionized
atomic species for which the potential of the upper level was smaller than the
ionization potential.  We also included the molecular lines of CN and of the
hydrides CH,  NH, and  OH.  For the hydrides we only considered lines with
excitation potential E.P $\leq $ 1.5 eV. In fact, since these molecules are
easily dissociated, the upper level for transitions starting at higher
excitational levels have energies that are much higher than the dissociation
energy, and are undetectable in both the solar spectrum and in our observed
spectra. We supplemented these lists with a few additional lines that are
listed in the solar tables (Moore,  Minnaert \& Houtgast,  1966) but are
missing from the Kurucz database. Most of these lines are unidentified
features; we arbitrarily assumed them to be Fe I lines with an E.P=3.5 eV. The
$\log (gf)$ of the atomic transitions from which we measured the abundances
were taken from the atomic physics literature (see the following sections for
details and references) and left untouched,  while the transition probabilities
of the neighboring lines were adjusted by matching the solar spectrum. A
synthetic spectrum of the wavelength range of interest was calculated using a
Kurucz model for the solar atmospheric parameters. The $\log (gf)$ were
adjusted to reproduce the observed solar spectrum,  taken from Kurucz's
(Kurucz {\it et al.} 1984) Solar flux atlas. In some cases small offsets (of
the order of 5$\times 10 ^{-3}$\,\AA) in the wavelengths were also applied.
With the exception of CH and CN, whose cases are discussed in \S 4.4.1 through
4.4.3, molecules were treated as contaminants. In order to reproduce the
broadening mechanisms (instrumental, macro-turbulence, rotation), the
calculated synthetic spectra were convolved with a Gaussian.  The FWHM for the
Gaussian were chosen according to the observed spectra. Once acceptable
agreement with the solar spectra was reached, the line list was used to
calculate the synthetic spectrum for the program star and thus to  measure the
abundance of the element of interest.

The observed data for \SarasStar\, were smoothed by convolving the {\it raw}
spectra with a Gaussian of FWHM = 0.1~\AA. Whenever possible, the observed
spectrum that was compared to the synthesis was the sum of the HIRES September
2000 and UVES August 2001 data. Whenever meaningful and possible,  we also
compared the spectrum of \SarasStar\, to that of \CompStar, taken in
September 2000, whose analysis was performed in Paper II. The two stars have
very similar atmospheric parameters (\SarasStar: 6625/4.3/$-$2.70/1.40;
\CompStar: 6625/4.3/$-$2.56/1.42),  so that the differences between the two
spectra can be attributed almost exclusively to the chemical peculiarity of
\SarasStar. In the text below we note when, because of the differences in the
wavelength coverage, only the UVES spectrum was used.

\subsubsection{C abundance \label{csyn}}

In order to measure the carbon abundance of \SarasStar, we calculated a
synthetic spectrum from 4300\,{\AA} to 4340\,\AA. This region includes the
band-head of the (0-0), (1-1), (2-2) bands of the A$^{2}\Delta$--X$^{2}\Pi$
transitions of CH. The dissociation potential 
used was 3.465\,eV from Brzozowski {\it et al.}(1976) and the $\log (gf)$'s 
of the electronic transitions were modified in
order to reproduce the solar spectrum using the solar abundance of carbon from
Anders \& Grevesse (1989) and the Kurucz solar spectrum. We found a corrective factor of 
-0.3\,dex in $\log g$ 
and a shift of -0.05\,\AA\,in wavelength.
%
%
%

\begin{figure}
\plotone{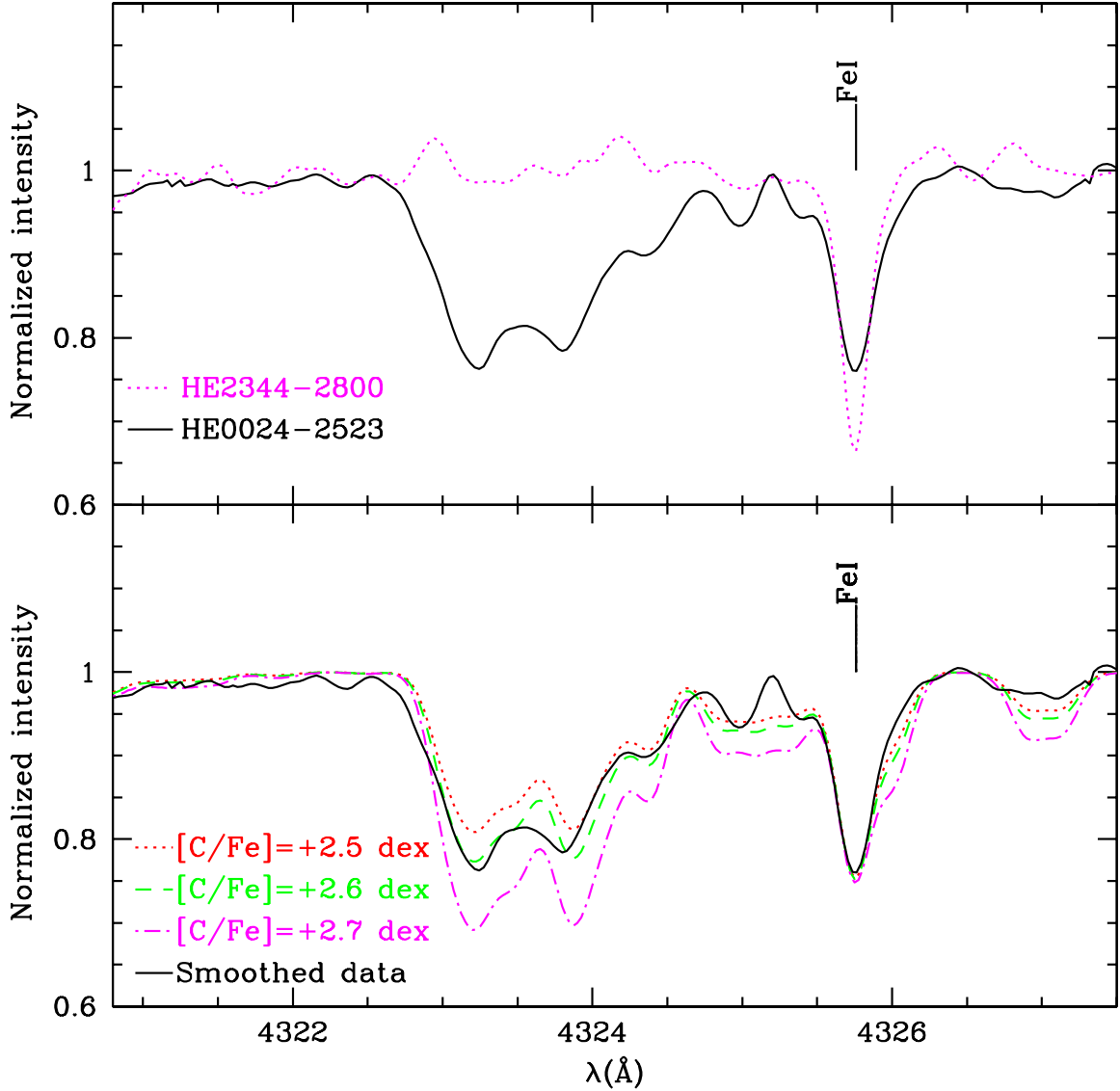}
\caption{ Comparison between observed and synthetized spectra for a spectral
region within the G-band. The continuous line is the observed spectrum. The
upper panel shows the comparison of the spectra of the program star
(continuous line) and of \CompStar\, (dashed line), which have almost
identical atmospheric parameters. The Q band-head which is undetectable in the
{\it normal} star,  is extremely strong in the program star. Note the
difference in the shape of the Fe I line at 4325.78\,\AA, due to the non
negligible rotational velocity of the program star ($\sim$ 9.7\,\kms). The
lower panel shows the comparison of the observed spectrum for \SarasStar\,
with three synthetic spectra computed with [C/Fe]=+2.5, +2.6 and +2.7\,dex.
\label{ch1} }
\end{figure}

We then computed the synthetic spectra with atmospheric parameters appropriate
for \SarasStar\, and carbon abundances from [C/Fe]=+2.0 to
[C/Fe]=+3.0, in intervals of 0.1\,dex. The lower panel of Figure
\ref{ch1}
shows the comparison between the observed spectrum and the synthetic
ones. For the sake of clarity we have only shown the synthetic spectra for the
cases [C/Fe]=+2.5, +2.6, and +2.7.  The overall best fit was achieved for
[C/Fe]=+2.6$\pm$0.1.  It is worth noting that the plotted synthetic spectra 
are the result of three separate steps.  We first measured the carbon abundance
using the solar isotopic ratio $^{12}$C/$^{13}$C;  then, assuming this $^{12}$C
abundance, we measured the isotopic ratio as described in the following section,
and finally we re-computed the synthetic spectra to determine the carbon
abundance using the measured isotopic ratio. The spectra with the correct
isotopic ratio exhibit a better fit to the data all over the entire range. The
upper panel of Figure \ref{ch1}
shows the comparison
between the spectrum of \SarasStar\, with the carbon-{\it normal} star
\CompStar, where CH lines are undetectable.

\subsubsection{C isotopic ratios \label{c13syn}}

\begin{figure}
\plotone{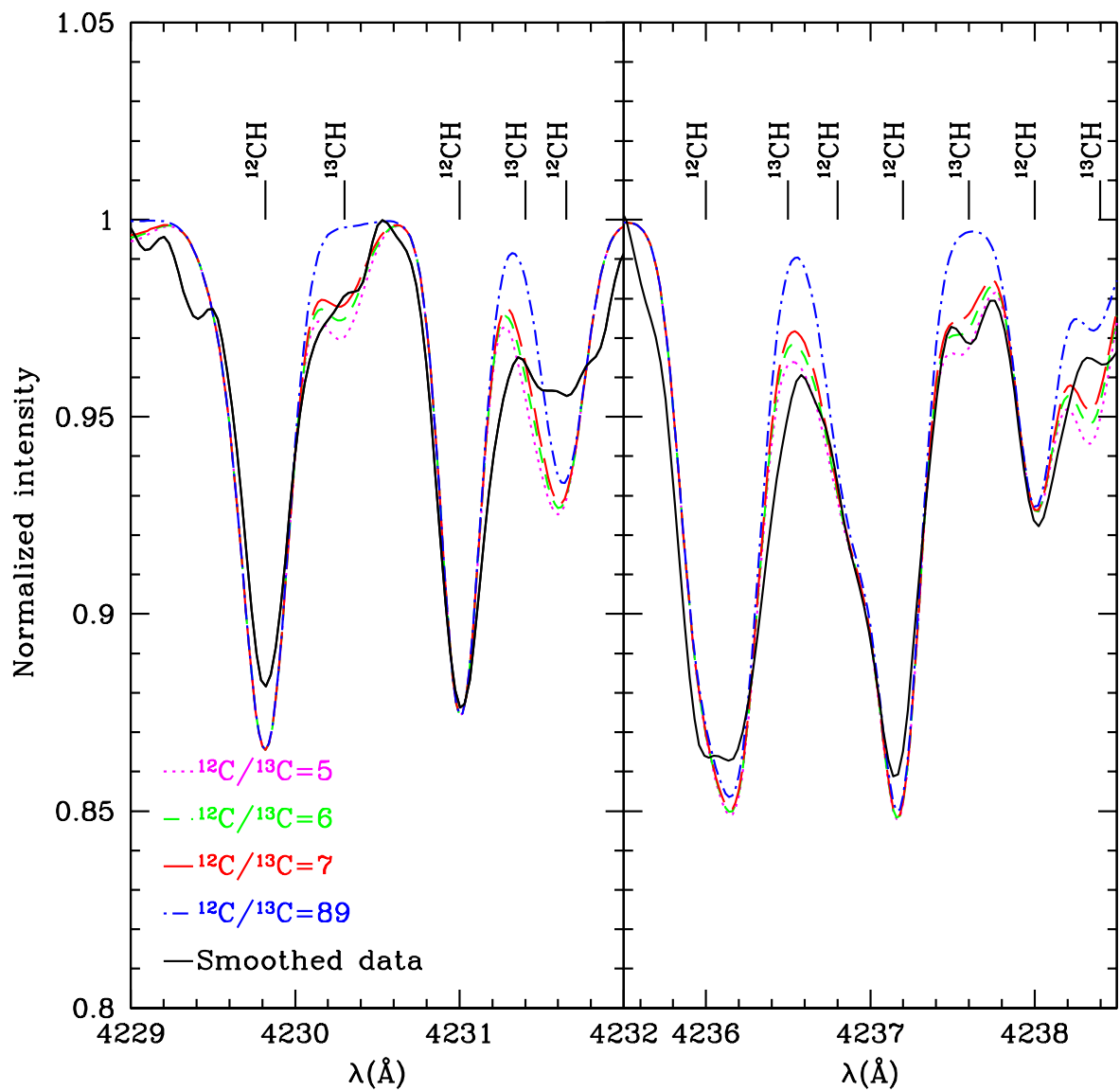}
\caption{ Comparison of observed and synthetized spectra in the regions 4229
to 4232\,{\AA}  ~and 4235.5 to 4239\,\AA. The solid line is the smoothed observed
spectrum. The synthetized spectra are for $^{12}$C/$^{13}$C=5,  6 and
7; a synthetic spectrum with  solar isotopic C ratio $^{12}$C/$^{13}$C=89 is
also plotted for comparison. Several $^{12}$CH and $^{13}$CH lines are showed
in order to highlight the features on which the measurement is based.
\label{ch13a}}
\end{figure}

\begin{figure}
\plotone{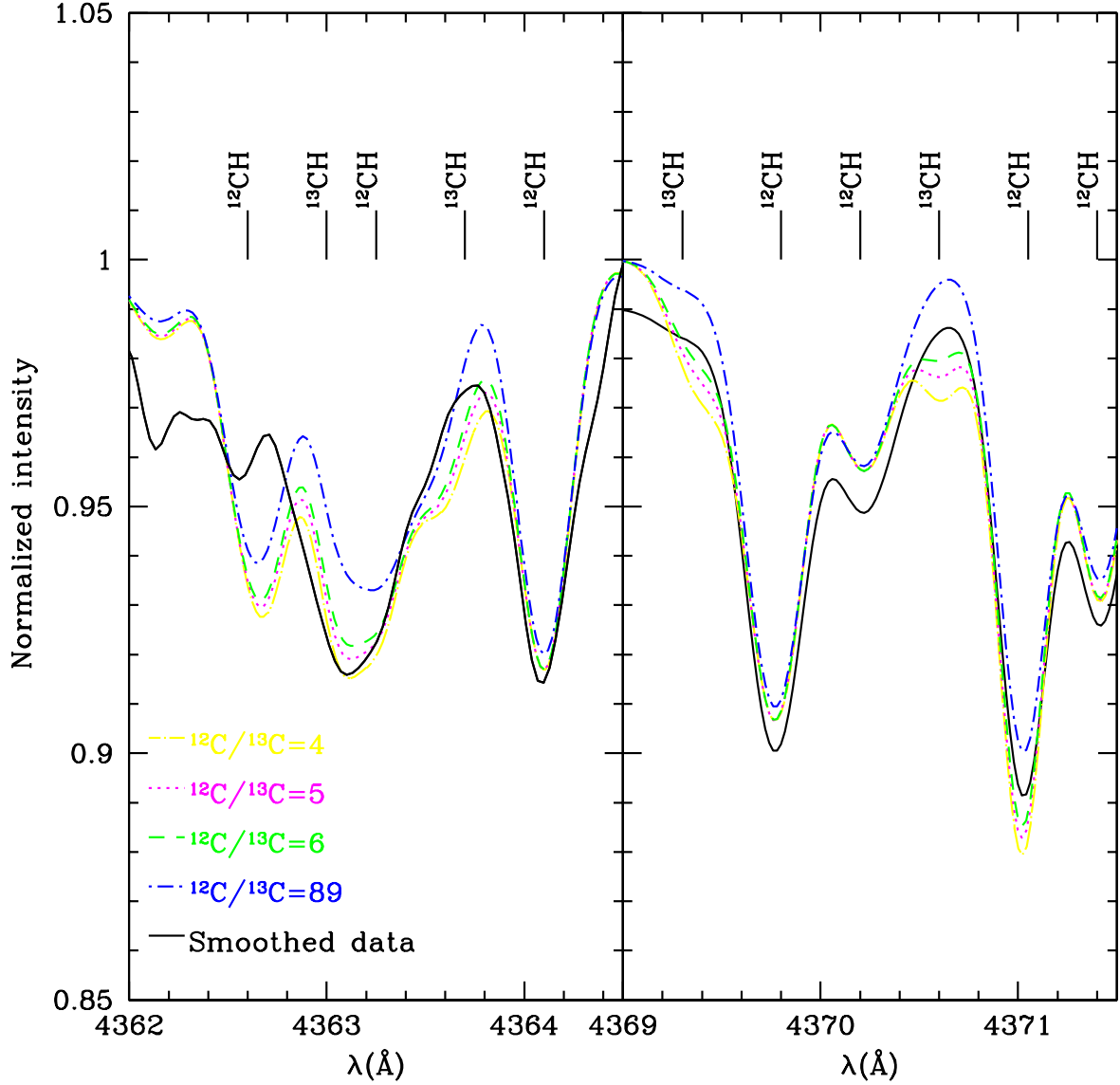}
\caption{As in Figure \ref{ch13a} but for different spectral ranges, (4362 to
4364.5\,{\AA} and 4369 to 4371.5\,\AA). The continuous line is the smoothed 
observed spectrum. The synthetized spectra are for $^{12}$C/$^{13}$C=4,  5 and
6; a spectrum with  solar isotopic C ratio $^{12}$C/$^{13}$C=89 is also
plotted for comparison. \label{ch13b}}
\end{figure}

The wavelength ranges of the spectrum used for the measurement of the isotopic
$^{12}$C/$^{13}$C ratio are 4229 to 4239\,{\AA} and 4362 to 4372\,\AA. The line
list was prepared as described in the previous section for the line list
associated with the G band.  We computed synthetic spectra for isotopic ratios
$^{12}$C/$^{13}$C from 3 to 10. The best fit for the region  4229 to
4239\,\AA\,was obtained for a value of $^{12}$C/$^{13}$C=6$\pm$1, while in the
region  4362 to 4372\,\AA\,we obtained $^{12}$C/$^{13}$C=5$\pm$1. We adopted
the ratio to be $^{12}$C/$^{13}$C=6$\pm$1 because in the first spectral range
the differences between the synthetic spectra due to the different assumed
isotopic ratios are clearly visible (see Figure \ref{ch13a}). In the
second spectral range the isotopic effects are more subtle, but the overall fit
is just as good (see Figure \ref{ch13b}).\\
%
%
%

\subsubsection{N abundance \label{nsyn}}

The N abundance was derived from the spectral region 3881-3885\,\AA, which
includes the band-head of the CN violet system. In this case the line list was
extracted from  Kurucz's database with the selection criteria and the
optimization procedure  described in \S \ref{synt}. We used a dissociation potential of 
7.66\,eV from Engleman \& Rouse (1975) and
even in this case 
we found a corrective factor of -0.3\,dex in the $\log (gf)$'s 
of the electronic transitions.
We computed synthetic
spectra using a [C/Fe]=+2.6, and with [N/Fe] abundances ranging from +1 to
+2.5 in 0.1\,dex increments. The best fit was obtained for  [N/Fe]=+2.1
$\pm $0.1.
%
%
%

\subsubsection{Ba and Eu abundance: {\it r} or {\it s}-process? \label{baeu}}

In our analysis we found a strong Ba overabundance (for a comparison with the
{\it normal } halo star, \CompStar,  see Figure \ref{ba}). In order to rule out
the possibility that Ba is due to the {\it r}-process,  we looked for other
prominent features of {\it r}-process elements. The most suitable element for
this test is Eu: its relative {\it r}-fraction in the solar system is about
93\% (K\"appeler {\it et al.},  1990a,  1990b). The $\log (gf)$ for the Eu II
line at 4129.72\,\AA\,was taken from Lawler {\it et al.} \cite{law}.  We also
took the hyperfine and isotopic splitting constants used to prepare the line
list from the same source, and adopted a solar-system isotopic ratio. As can be
seen from Figure \ref{eu},  there is no clear evidence of a line at
4129.72\,\AA. An upper limit to the Eu abundance of [Eu/Fe]
$\leq+1.1\pm0.4$ is deduced.
%
%
%

While this limit is not known with great accuracy,  an {\it r}-process origin
of Ba can be excluded. In fact,  the expected Eu abundance in the case of {\it
r}-process enrichment would be [Eu/Ba]$\sim$0.8,  that is to say, for
[Ba/Fe]=+1.46, we would expect [Eu/Fe]$\simeq$+2.2. This Eu
abundance is not consistent with the observed data, as Figure \ref{eu} clearly
shows. The dotted-dashed line, which represents the synthetic spectrum for
[Eu/Fe]=+2.2,  is characterized by an Eu feature  much stronger than that
present in the spectrum of \SarasStar, even with a generous allowance for errors in 
the placement of the continuum.

\subsubsection{Pb abundance \label{pbsyn}}

Having ruled out the {\it r}-process origin of the Ba in \SarasStar, we
expect the Ba to originate from the {\it s}-process. According to most
nucleosynthesis models, the flow of the {\it s}-process at low metallicities,
given the scarcity of seed nuclei, drains the Zr and Ba peaks and builds an
excess of $^{208}$Pb,  which is at the end of the {\it s}-process path (Busso
{\it et al.} 2001).  We measured the Pb abundance from two of the
strongest lines in our spectral range: 3683.4 and 4057.8\,\AA. The transition
probabilities for the two lines were taken from Hauge \& S{\o}rli
\cite{hau}\footnote{These $\log (gf)$ are those used by Anders \& Grevesse
\cite{and},  whose solar abundances are adopted in our analysis and synthesis.
They find log $\epsilon$(Pb)=+1.85$\pm$0.05. More recent values of $\log
(gf)$ (Bi{\' e}mont {\it et al.} 2000) for those two lines are smaller by about
0.11\,dex, and the solar abundance obtained is correspondingly higher,  log
$\epsilon$(Pb)=+2.00$\pm$0.06.   The two effects essentially cancel out, and
given that we use the old $\log (gf)$ and old solar lead abundance, the [Pb/Fe]
abundance derived for the program star is not significantly affected.}.

Neither hyperfine structure nor isotopic splitting was included;  the
inclusion of these effects, which is probably worth doing when spectra of
higher resolution and signal-to-noise ratios are available, would tend to
decrease the measured Pb  abundance. However, desaturation effect of the
hyperfine splitting should not be very important in these lines, which are
quite weak. Figure \ref{pb1} shows the spectral region around the 3683.5\,\AA~
line together with three synthetic spectra calculated for  [Pb/Fe]=+3.2, +3.4,
and +3.6, respectively. To illustrate just how strong the Pb line is, we have
plotted a spectrum with a scaled solar lead content. The lead abundance
measured from the 3683.4\,{\AA} line is [Pb/Fe]$=+3.4\pm0.2$.  No comparison
with \CompStar\, is shown as this line is outside the spectral range covered by
our HIRES spectra.
%
%
%

The spectral region around the 4057.7\,{\AA} line is covered by both UVES and
HIRES spectra, so the signal-to-noise ratio is higher. Figure \ref{pb2}
shows,  in the lower panel,  the observed spectra and the
synthetic spectra computed for lead abundances [Pb/Fe]=+3.1, +3.2, +3.3,
respectively, as well as the scaled solar value.  From this line we measured a
lead abundance of [Pb/Fe]=+3.2$\pm$0.1. The comparison with the spectrum
of \CompStar, in the upper panel of Figure \ref{pb2},  shows a large feature in
the spectrum of \SarasStar\, at 4057 \AA\, due to the lines of  $^{12}$CH, Fe I
and Pb I, while only the Fe I line can be seen in the {\it normal} spectrum.
Also, the blend of the two $^{12}$CH lines at $\sim$ 4059.4\,{\AA} is barely
undetectable in that spectrum. We adopt the lead abundance for the program
star from the mean of the two determinations: [Pb/Fe]=+3.3$\pm$0.1.  This value
might be slightly modified if the  hyperfine and isotopic splitting were
included. As mentioned,  since the lines are quite weak, this would not lower
the measured abundance by more than about 0.1\,dex.
%
%
%
%

Thus, even taking the above factors into account, this is the most extreme {\it
s}-process enrichment yet measured in a extremely metal-poor star. While the
Pb abundance is extraordinarily high, the ratio [Pb/Ba]=+1.86$\pm$0.22 ,
which represents the ratio of abundances between elements at the second and
third {\it s}-process peaks, is the highest ever measured (see Table
\ref{chs}). The lead content in the Sun is about 1 Pb atom for every 3$\times
10^{5}$ Fe atoms, while in this star it is about 1 in 100. The previous record
belonged to CS\,31062-059 (Aoki {\it et al.} 2002c) which has a metallicity
[Fe/H]=$-$2.38\,dex, a lead content identical to that we have measured for
\SarasStar\ ([Pb/Fe]=+3.3$\pm$0.24), but a ratio [Pb/Ba]=+1.19$\pm$0.14 .
%
%
%
\subsubsection{La abundance \label{lasyn}}

We measured the lanthanum abundance from the 3988.5 and 4333.8\,{\AA} lines of
La II. The $\log (gf)$\ and the hyperfine splitting constant for these lines were
taken from Lawler, Bonvallet \& Sneden \cite{law2}. We did not include isotopic
shifts, since in the Sun the isotope $^{159}$La accounts for more than 99.9\%
of the total La content. The abundance measured from both lines is [La/Fe]=+1.8
$\pm $ 0.2 (see Figures \ref{la1} and \ref{la2}). The error is quite large
because the signal-to-noise in the spectral region around both lines is not
very good.
%
%
%

\subsubsection{Li abundance \label{lisyn}}

We measured the Li abundance of \SarasStar\, using the 6707\,{\AA} resonance
line.  We inspected this element not because it is expected to partake in {\it
s}-process nucleosynthesis, but because of recent results reported by Ryan {\it
et al.} (2002).  These authors, on the basis of the analysis of a sample of 18
halo main-sequence turn-off  stars, suggest that the Li abundance in their
sample seems to be anti-correlated with rotational velocity.
 
Figure \ref{li} displays the observed and synthetized spectra in the region of
the Li doublet; no definite feature was detected at 6707\,\AA, hence we deduce
an upper limit of $\log \epsilon$(Li)=+1.4.
%
%
%

\subsection{Abundance summary \label{abusum}}

Table \ref{abu} shows a summary of measured elemental abundances for
\SarasStar. The last column indicates whether the listed abundance was derived
via a line-by-line analysis of measured equivalent widths (indicated with
``EWs'') or with spectral synthesis (``syn''). For the abundance obtained from
the analysis of equivalent widths, the r.m.s. is the dispersion of abundances
as measured from the different lines, while the listed errors take into
account the scatter in the abundances derived from the lines  ({\it i.e.} the
r.m.s. divided by the square root of the number of the lies) and the effect of
the uncertainties in the atmospheric parameters, as explained in \S
\ref{synt}. Also reported are the C isotopic ratio, and the ratios C/O and N
/O, by number.

\section{Discussion \label{disc}}

Our results clearly indicate that the chemical composition of the atmosphere
of \SarasStar\, is the outcome of an {\it s}-process enrichment. The abundance
pattern observed is very extreme, especially as far as Pb is concerned.
Canonical {\it s}-process nucleosynthesis is associated with the AGB stage of
evolution, during which the star is powered by hydrogen and helium burning in
two shells located above a degenerate core (see,  {\it e.g.},  Iben \& Renzini
1983). Late on the AGB,  intermediate mass stars undergo recurrent thermal
instabilities of the He shell, in which partial He burning occurs convectively
over short periods of time,  sweeping across the whole region that lies between
the H and the He shells. During this phase,  the material from the He
inter-shell  that is enriched in C and {\it s}-process elements is taken to the
surface by the extension of the envelope convection (third dredge up).  Even
later in the evolution, as the result of stellar winds which progressively
strip its envelope,  only the degenerate core is left and the star will become
a white dwarf (see Busso {\it et al.}, 2001, for a thorough discussion of {\it
s}-process elements nucleosynthesis).

As discussed in Paper I, the  observational data indicate that \SarasStar\, is
presently going through its upper main-sequence phase and thus is not capable
of having produced the abundance pattern we have measured. The necessity of an
external origin of some of the detected elements, and the
clear binarity of the object, suggests that \SarasStar\,is an {\it extrinsic}
CH star, {\it i.e.} the atmosphere of the star observed has been polluted by
material that has been processed by a more massive companion which already
underwent its AGB phase and is now a white dwarf (and undetectable in our
present observations).  This model is further supported by consideration of the
orbital solution, as discussed in \S 3.3 above.

\subsection{Orbital parameters \label{orbdisc}}

It is noteworthy that the orbital period of \SarasStar\, is unusually short,
compared with those typical of this class of object, which is several hundred
days (McClure \& Woodsworth, 1990). This anomaly suggests that \SarasStar\,
went through a common envelope phase at an earlier stage of the binary
interaction, when the more massive component expands so rapidly that mass
transfer is unstable, and the companion is engulfed in the atmosphere of the
donor star. The ejection of the envelope requires energy to be removed from the
orbit, and the latter shrinks, shortening its period (see, {\it e.g.},
Sandquist {\it et al.}, 1998).

This picture could explain the peculiarity of this star as the result of a more
pronounced difference between the original masses of the components of the
binary, since the common envelope phase is the consequence of unstable mass
transfer. This is consistent with the measured C isotopic ratio, which, as
discussed later in this section, suggests an initial mass for the evolved
companion of at least $\sim$3\,M$_{\odot}$. Had the stars been more similar in
mass, the mass transfer might have taken place quiescently, and the orbit would
have shrunk much less, probably reaching the typical CH star period.
 
\subsection{Testing theoretical models \label{theordisc}}

We now compare our measured abundances for the {\it s}-process elements to
those predicted from an evolution model computed for a metallicity of
[Fe/H]=$-$2.6 and mass M=1.5\,M$_{\odot}$ (Gallino 2002,  priv. comm.),
which we consider here as representative of the now evolved-companion of
the observed star. In order to compare the predictions with the observations,
we assumed a {\it dilution factor}, the percentage of
atmospheric material which has been accreted by the observed star from the
donor star, of 10\%. This assumption, while reasonable, is completely
arbitrary, and was set to match the observed carbon abundance. We also made the
natural assumption that the original ({\it i.e.} pre-accretion)
abundance pattern of the observed star resembled that of a typical star of
similar metallicity. For this purpose, we used the mean of the abundances
measured in several stars at [Fe/H]$\sim-$2.7 (see \S 5.2 of Paper II
for more details on the sample).

The comparison is shown in Figure \ref{theory}. Several elements shown
have negligible contribution from the shell nucleosynthesis (Mg
through Fe), thus the fit is not very meaningful for them. The prediction in
the case of Pb and Ba, produced exclusively by AGB nucleosynthesis, is
excellent, with the model falling well within the error bars. On the other
hand,  Sr is not reproduced particularly well.  However, it should be kept in
mind that the contribution of the original (pre-accretion) Sr content to the
final one has to be considered as well; the measured Sr abundance in
metal-poor stars is known to have a large scatter. Thus, it is possible that
the average Sr content that we assumed as representing the original abundance
of this element does not pertain.  The measured upper limits for the abundances
of Eu, La, Y and Zr are indicated by arrows. All of them are consistent with
the model.
%
%
%

While the model seems to reproduce satisfactorily our measurements for {\it
s}-process elements, it fails to  explain the low $^{12}$C/$^{13}$C ratio,
predicting a value of several hundred (Gallino 2002,  priv. comm.). The low
measured ratio ($^{12}$C/$^{13}$C = $6 \pm 1$) is quite unusual for {\it
s}-process enriched material. In fact, the {\it primary} source of neutrons is
thought to be the $^{13}$C($\alpha$, n)$^{16}$O reaction, hence $^{13}$C is
expected to be highly depleted.  The value of the isotopic ratio is only
marginally larger than that of CN cycle equilibrium value
($^{12}$C/$^{13}$C$\sim3$), and seems to suggest a signature of  hot bottom
burning (HBB hereafter).
 Since the observed star is still on the main sequence,  it has not
gone through the first dredge-up and all the carbon observed, both $^{13}$C and
$^{12}$C, presumably  comes from the former AGB donor star.

On the basis of the hypothesis that the donor star underwent HBB before mass
transfer took place, we can place a reasonable lower limit to the {\it
original} mass of the donor star, which is the minimum mass for which HBB is
expected to occur.  This is $\sim 3{\rm M}_{\odot}$ (Marigo 2002,  private
communication) for the metallicity under discussion. This value is lower than
the usual one, $\sim4.5\,{\rm M}_{\odot}$ (Marigo 1998), because, as
metallicity declines, the mass at which the temperature necessary for igniting
HBB is reached also decreases (see Siess,  Livio,  \& Lattanzio 2002 for an
extreme case at zero metallicity). It should be noted that the initial mass for
the shell nucleosynthesis model we used for comparison, M=1.5\,M$_{\odot}$,  is
smaller than the lower limit placed when taking into account the low
$^{12}$C/$^{13}$C.  HBB, which would lower the above mentioned ratio,  possibly
approaching that of our measurement, is not achieved in the model.

A recent paper by Ventura, D'Antona, \& Mazzitelli (2002) presents the yields
from AGB stars with $2.5 \le \mbox{M}/\mbox{M}_{\odot}\le 6$ and
$-2.0 \le \mbox{[Fe/H]} \le -0.3$.  Their results seem to agree with
our findings for \SarasStar, both for the isotopic C ratio and C/O ratio.
Nevertheless,  the N abundance and the N/O ratio are  not reproduced in these
models,  which predict that the N abundance  should be about an order of
magnitude larger than that of C. As shown in Fig \ref{theory}, unlike the
data in the literature, the extreme value of [Pb/Ba] for \SarasStar\, is
very well explained by models with the standard mass of the $^{13}$C pocket
(Busso {\it et al.} 2001), which predict [Pb/Ba]$\sim$2.  On the other hand,
the values found in the literature (see Table \ref{chs}) are lower and, in
order to be matched by the models, would require a reduction of the mass of
the $^{13}$C pocket by factors 6 to 24 (Aoki {\it et al.} 2002c). 
Although the data for the $^{12}$C/$^{13}$C ratio in other similar stars are
still very limited, it is noteworthy that the value for \SarasStar\, is the
lowest of the sample, consistent with the suggestion that the donor star had
an unusually high initial mass, as discussed above.

\subsection{Comparison with literature data for {\it s}-process enriched, extremely 
metal poor stars \label{complit}}

Figure \ref{comp} displays the ratio [Pb/Ba] for \SarasStar\, and for
data from the literature (Aoki {\it et al.} 2000; Aoki {\it et al.} 2002b,c)
as a function of [Fe/H]. Aoki {\it et al.} (2002c) suggested that the
available data might show evidence of a decreasing trend of [Pb/Ba] with
decreasing [Fe/H]. Our analysis seems
not to support this hypothesis, at least in all cases. It is quite clear from
Fig \ref{comp} that, once the data point for \SarasStar\, is added, no
correlation seems to be present.  However, caution should be exerted when
drawing any conclusion from this apparent lack of correlation. \SarasStar\, is
the only star in the sample which likely went through a common envelope phase,
which may well have influenced the nucleosynthesis history of the donor star.
Thus, the exceptionally high [Pb/Ba] for such a low metallicity
([Fe/H]=$-$2.7), with respect to those found in the literature, could arise
from the peculiar evolution of \SarasStar, and not destroy the
[Pb/Ba] vs [Fe/H] correlation suggested by Aoki {\it et al.} (2002c).
Presumably, the addition of more stars with measured [Pb/Ba], especially at the
lowest metallicity, will help clarify the situation. 
%
%
%

Another noteworthy feature seen in Figure \ref{comp} is the larger scatter in
[Pb/Ba] about the regression line for stars with detected radial
velocity variations.  For these objects, the observed {\it s}-process
enrichment is a signature of the nucleosynthesis which took place on the
companions during their AGB phase and was then accreted by the observed stars.
Thus, our results indicate that the ratio [Pb/Ba] produced by the {\it
s}-process in metal-poor stars ([Fe/H]$<-1.9$) shows a large scatter.

A possible explanation for this scatter could be a correlation between the
[Pb/Ba] ratio and the radial velocity variations. In fact, we could invoke the
presence of a close companion as a factor affecting shell nucleosynthesis. If
the orbital period is large, the donor star will undergo complete evolution
along the AGB, and the composition of the accreting star will be dominated by
material lost by the evolved star just before the final thermal pulse and
planetary nebula ejection. On the other hand, a very short period, like the
one observed in the case of \SarasStar, indicates a past common-envelope
phase; during this stage, a large amount of mass is lost by the system,
leading to a premature end of the thermal pulse phase of the donor. In this
case the chemical composition of the accreting star will be dominated by
material lost by the evolved star during earlier phases of the evolution along
the AGB. This may have a big impact on HBB processes, affecting the abundances
of $^{12}$C, $^{13}$C, N and possibly even O, and thus the entire {\it
s}-process history, which relies on $^{13}$C pocket as a neutron source.

\subsection{\SarasStar: a future blue straggler? \label{bss}}

A number of recent papers ({\it e.g.}, Preston \& Sneden 2000; Carney {\it et
al.} 2001; Ryan {\it et al.} 2001), considering a wide range of evidence ({\it
e.g.}, the observed high binary fraction amongst field blue stragglers, the
distribution of observed orbital periods and eccentricities in metal-poor
carbon-enhanced stars), have come to the conclusion that the formation of blue
stragglers is closely connected to the presence (now, or in the past) of a
companion star which has undergone mass transfer.  Ryan {\it et al.} (2001), in
particular, have argued that Li depletion in mass-transfer binaries (even those
which are presently too cool to be referred to as blue stragglers) might
account for the lack of Li observed in the field blue stragglers that they have
either already evolved into, or which they will eventually evolve into.  Ryan
{\it et al.} (2002) have recently provided clear evidence that a large fraction
of halo main-sequence stars that exhibit extreme Li depletion {\it also} have
substantial axial rotation, with $vsini$ in the range $5 - 8$\,\kms.  One is
thus tempted to consider the observed properties of \SarasStar, and ask whether
it might present an extreme example of these phenomena.

Clearly, the presence of carbon-enhancement and {\it s}-process elements in 
\SarasStar\, are consistent with previous episodes of mass transfer.  The
observed upper limit of the lithium abundance, ($\log \epsilon$(Li) = +1.4, see
section \ref{lisyn}), is quite in line with the observed upper limits of the
Li-depleted stars discussed by Ryan {\it et al.} (2001) (see their Figure 2).
The estimated $vsini$ of \SarasStar\, is at the high end (though still
consistent with) the range of the Ryan {\it et al.} (2002) stars.  The
orbital eccentricities of the Ryan {\it et al.} (2002) Li-depleted stars are
low as well, $e \le 0.3$ (one, BD$+51^{\circ}1817$, is very close to circular),
as are the eccentricities of the field blue stragglers of Carney {\it et al.}
(2001).  In the case of the latter, $<e> = 0.11$.  It is worth noting that
several of the blue stragglers in the Carney {\it et al.} (2001) sample exhibit
detectable axial rotation, on the order of 9-10\,\kms, similar to what we
observe in \SarasStar.  As has been emphasized by Ryan {\it et al.} (2002),
such high rotational velocities would not be naively expected for (presumably)
quite ancient stars of the halo population, and must have resulted from some
late-stage spin-up, quite likely at the time of mass transfer.

The one observable that clearly distinguishes \SarasStar\, from the stars
studied by Ryan {\it et al.} (2002) is its extremely short orbital period,
$~$3.4 days, as compared to the typical orbital period of 200-700 days in the
latter sample.  However, there are three field blue stragglers in the Carney
{\it et al.} sample, CS~22170-028, CS~22873-139, and CS~22890-069, which
possess orbital periods of from 1 to 20 days, similar to \SarasStar\, (the rest
of the Carney et. al.  sample exhibit orbital periods that are substantially
longer).  The axial rotations of these three stars have not been measured.  The
case of CS 22873-139 is particularly interesting, as this star was shown to be
a double-line spectroscopic binary by Preston (1994), who argued that at least
the primary of this system might be identified with the ``blue metal-poor
main-sequence'' (BMP) population of halo stars noted by Preston, Beers, \&
Shectman (1994).  These BMP stars are the very same population which Preston \&
Sneden (2000) showed to be dominated by binaries, presumably many of which
either are, or will become, field blue stragglers.  In a detailed
high-resolution study by Spite {\it et al.} (2000), this star was shown to
exhibit an unusual pattern of elemental abundances, with strikingly low [Sr/Fe]
($< -1.1$), [Mg/Fe] ($-0.04$), [Ca/Fe] (+0.16), and [Al/Fe]
($-$0.80), both compared to most other halo stars, and to \SarasStar.
There is no evidence of carbon enhancement in this star.  The Li doublet at
6707\,\AA\, was not detected, though Spite {\it et al.} argued that the
abundance of lithium might still be consistent with the primordial value, based
on the quality of their available data.  Clearly, the abundance variations
among the populations of stars that might be associated with the blue-straggler
phenomenon has yet to be fully explored. 

\section{Conclusions}

The peculiarity of \SarasStar, and the comparison of its measured abundances
and orbital parameters with other metal-poor stars in the literature suggest a
correlation between binarity and observed {\it s}-process element patterns.
This hypothesis is based on a very limited number of objects, and a larger
sample of well-studied CEMP stars is required to improve our knowledge of the
different evolutionary states, formation scenarios, and nucleosynthesis
histories of these objects. To this end, we have already obtained spectra with
HIRES of four more examples of apparent CEMP stars selected in the same
way as described in Paper I of this series.  We are aware of a number of other
groups that are pursuing related observational campaigns, so the database of
available information is expected to expand in the near future. 

\acknowledgments
S.L., R.G and E.C. acknowledge partial support from the MURST COFIN 2000.
J.G.C and S.R. are grateful for partial support from the Fullham Award of the
Dudley Observatory, and from grants AST 98-19614 and AST 02-05951 awarded by
the US National Science Foundation. N.C. acknowledges financial support
through a Marie Curie Fellowship of the European Community program
\emph{Improving Human Research Potential and the Socio-Economic Knowledge}
under contract number HPMF-CT-2001-01437, and from Deutsche
Forschungsgemeinschaft under grant Re~353/44-1. T.C.B. acknowledges partial
support for this work from grants AST 00-98508 and AST 00-98549  awarded by the
US National Science Foundation.

The authors are grateful to R.~E.~M. Griffin for supplying the {\it
Orbitsolver} code.  They also thank P. Bonifacio for performing part of the
observations, and R. Gallino, F. D'Antona, P. Marigo and O. Straniero
 for useful discussions.

\clearpage

\clearpage

\begin{figure}
\plotone{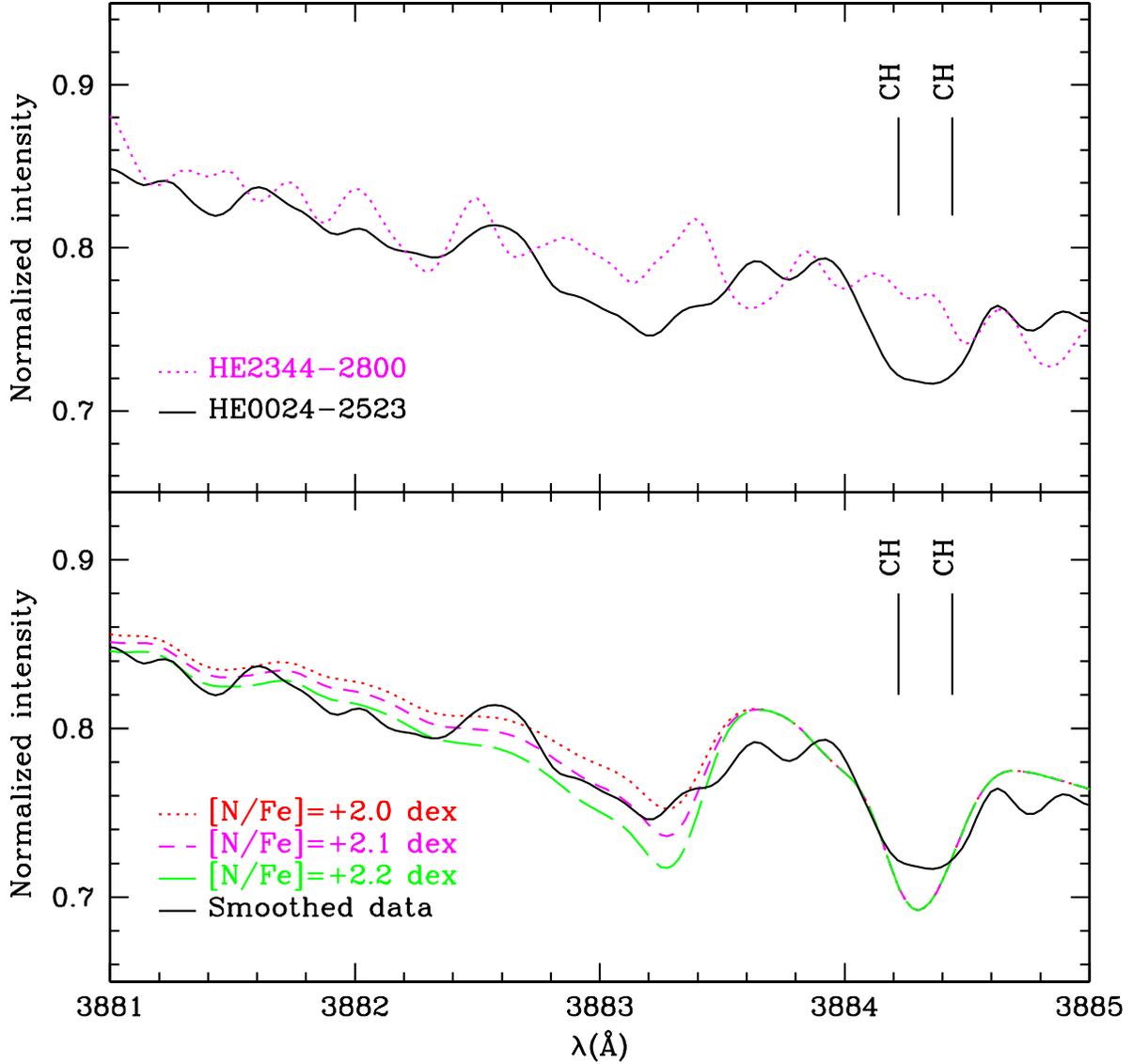}
\caption{The lower panel shows the comparison of observed and synthetized
spectra in the region of the UV CN head-band. The solid line represents the
smoothed observed spectrum. The synthetic spectra plotted are those for
[N/Fe]=+2.0 +2.1 and +2.2\,dex. The continuum is heavily affected by the wings
of the Balmer line H$_{8}$ at 3889.9\,\AA. In the upper panel, the program
star spectrum is plotted,  in the same spectral range,  together with the
spectrum of our comparison star,  \CompStar\, which does not show evidence of
CN band features. Note also the CH feature at $\sim$3884.3\,{\AA} is completely
lacking in the spectrum of \CompStar. \label{cn}}
\end{figure}

\begin{figure}
\plotone{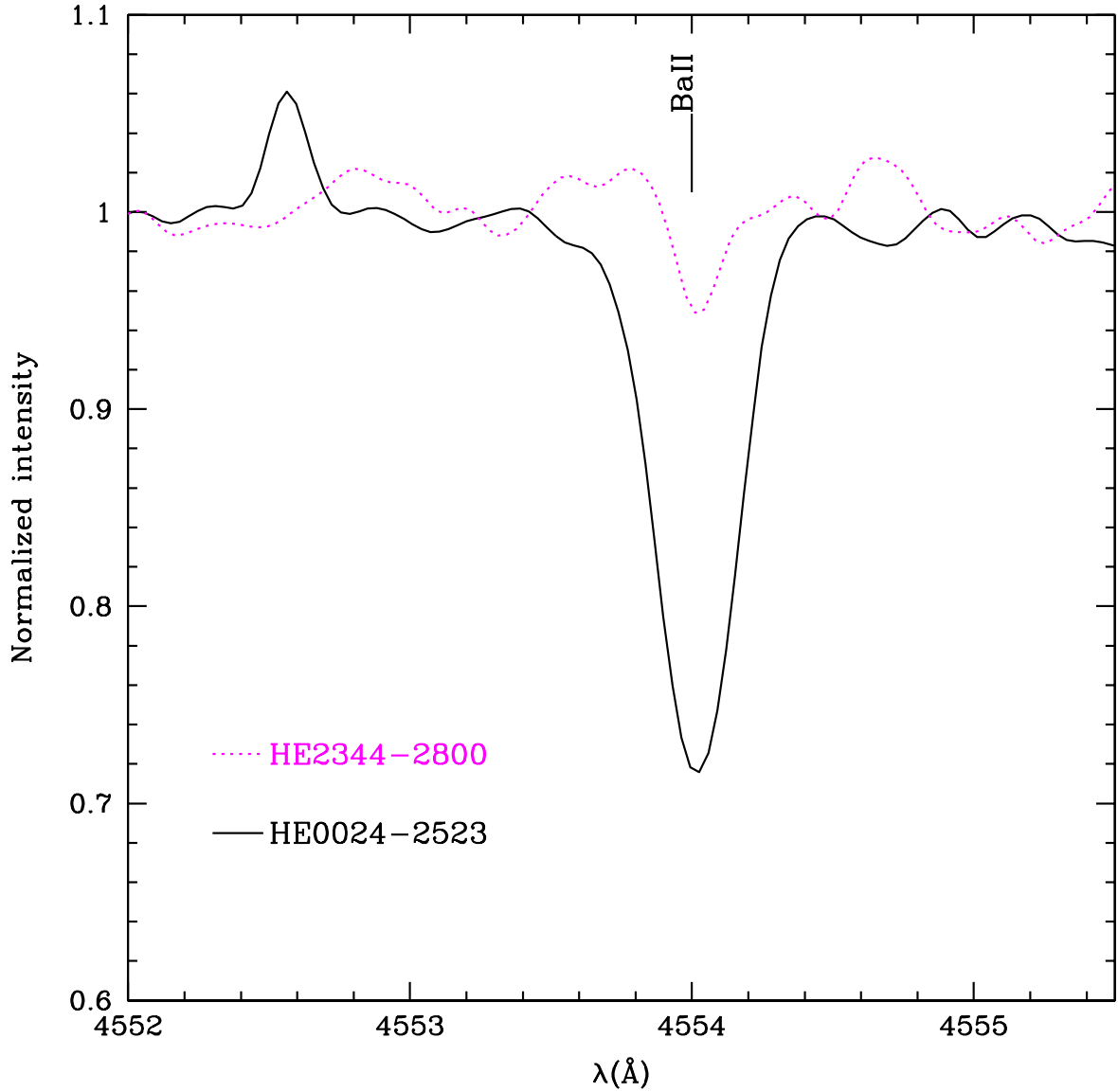}
\caption{Comparison between the smoothed spectra of \SarasStar\, and of
\CompStar\, in the spectral region around the Ba II 4554\,{\AA} line. The
difference is remarkable and reflects the measured Ba abundances:
[Ba/Fe]II=+1.46$\pm$0.2\,dex in \SarasStar\, (see section \ref{synt} for
details) and [Ba/Fe]II=$-$0.55$\pm$0.2\,dex for \CompStar. \label{ba}}
\end{figure}

\begin{figure}
\plotone{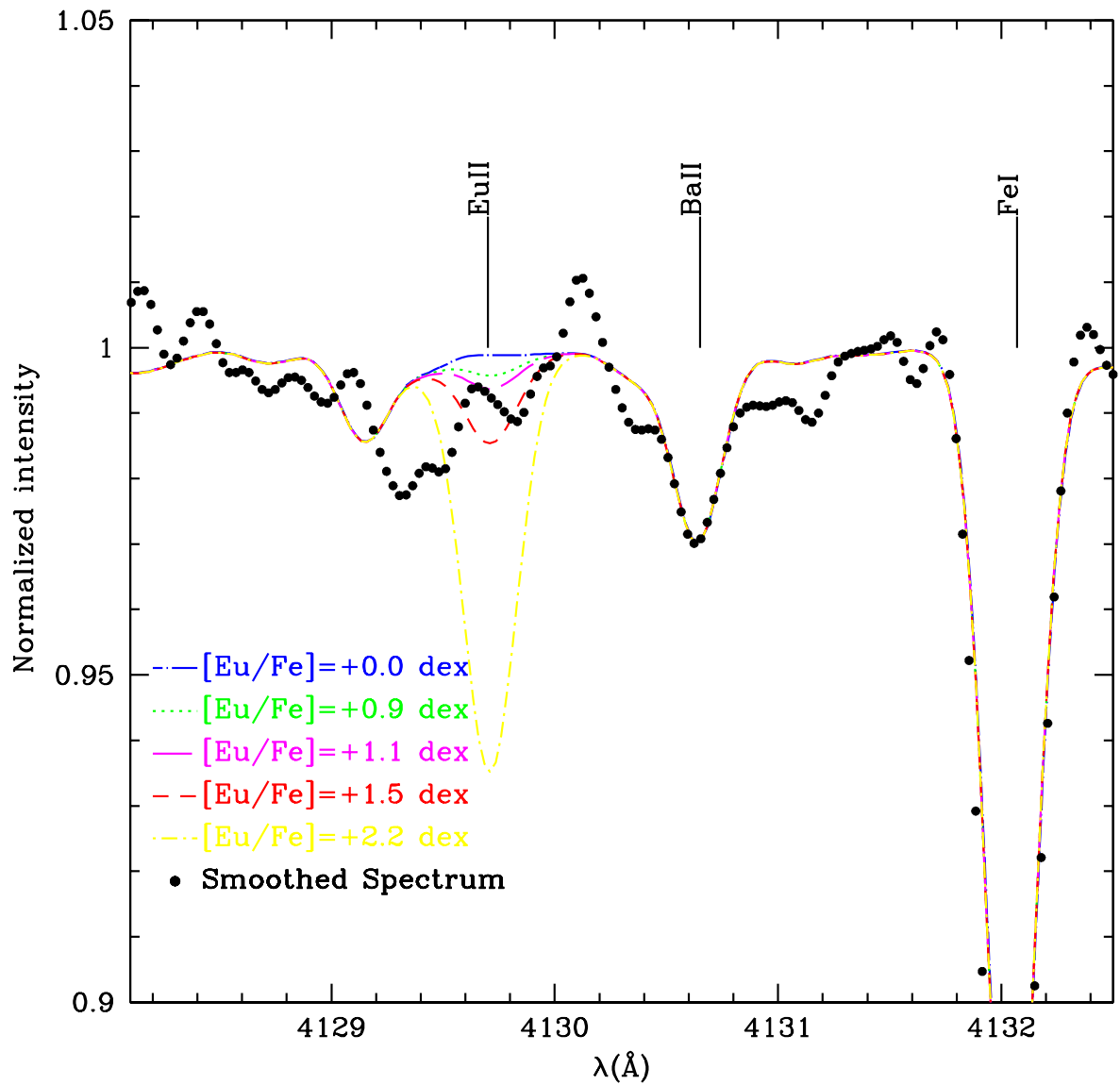}
\caption{ Comparison of the observed spectrum with synthetized ones for the Eu
II line at 4129.72\,\AA. The dots represent the smoothed observed spectrum.
The synthetic spectra were computed for [Eu/Fe]=+0.9,  +1.1 and +1.5\,dex.
Also shown are a spectrum calculated with solar [Eu/Fe] and with
[Eu/Fe]=+2.2\,dex which, given the measured Ba abundance of [Ba/Fe]=+1.46\,dex, 
represents the expected Eu abundance in case of {\it r}-process elements
enrichment. \label{eu}}
\end{figure}

\begin{figure}
\plotone{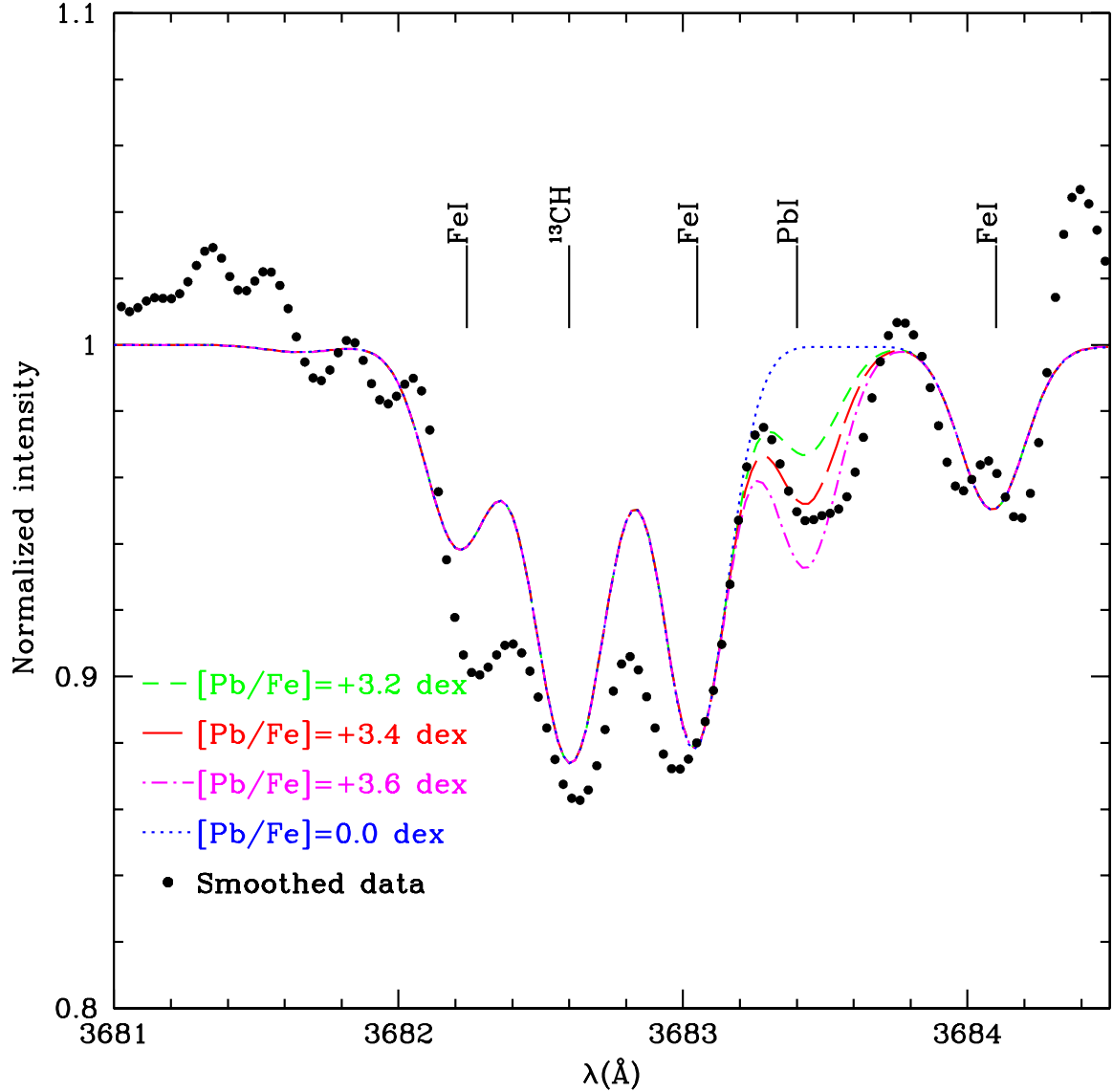}
\caption{Comparison of the observed spectrum with the synthetic spectra in the
spectral region around the Pb line at 3683\,\AA. The smoothed observed
spectrum is represented by the dots. Synthetic spectra plotted are for lead
scaled solar abundance,  [Pb/Fe]=+3.2, +3.4 and +3.6\,dex. The spectrum
appears very noisy because it is at the blue end of the region observed and 
only data from UVES were available, as this line is outside the wavelength
range covered by our HIRES spectra. \label{pb1}}
\end{figure}

\begin{figure}
\plotone{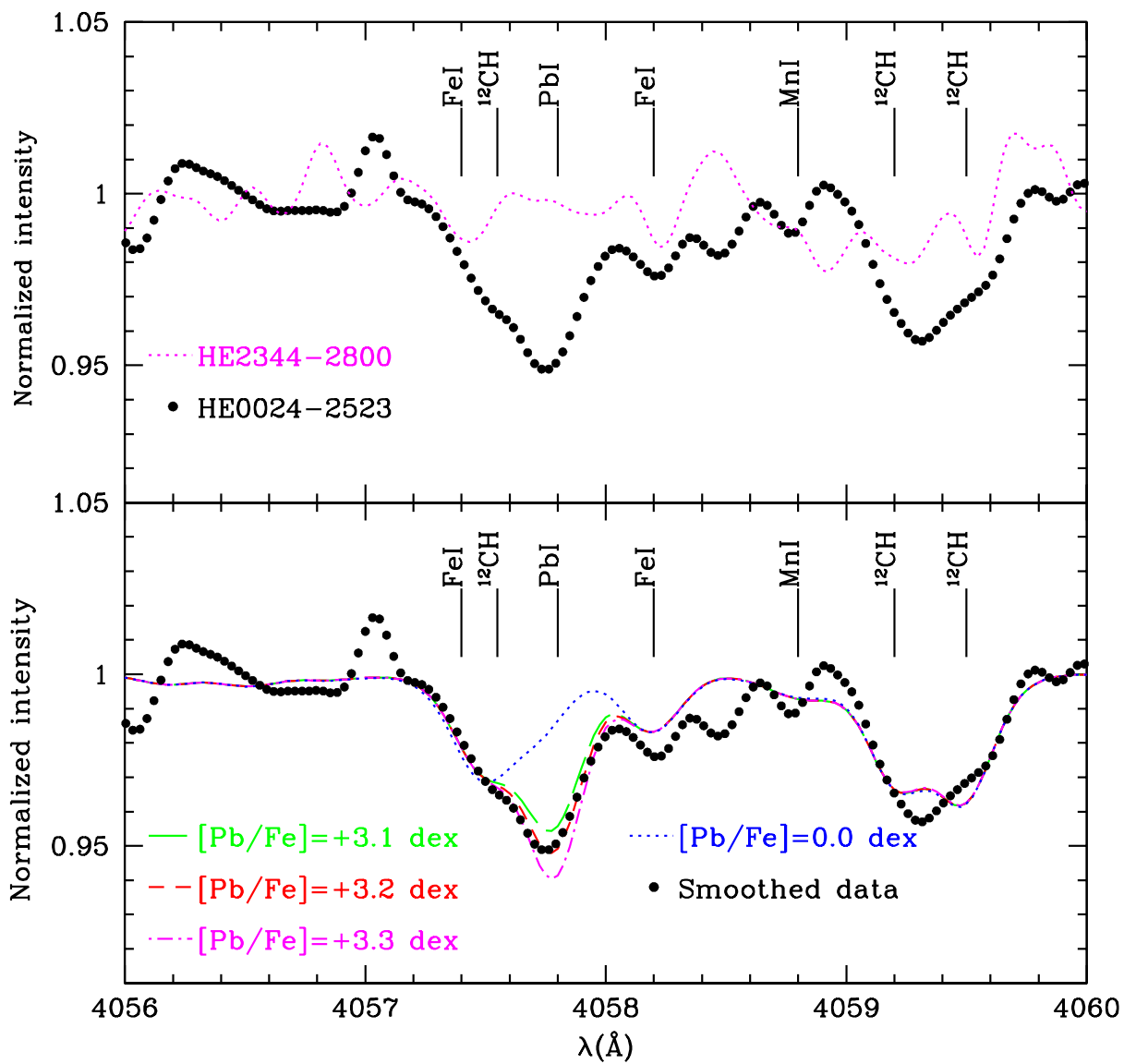}
\caption{Lower panel: comparison of the smoothed observed spectrum with the
synthetic ones in the region around the Pb line at 4057\,\AA. Synthetic
spectra plotted are for lead scaled solar abundance,  [Pb/Fe]=+3.1, +3.2 and
+3.3\,dex. Upper panel: comparison with the spectrum of the normal abundance
pattern star \CompStar. Notice the difference in the Pb I 4057.7\,{\AA} line
and in the CH features. \label{pb2}}
\end{figure}

\begin{figure}
\plotone{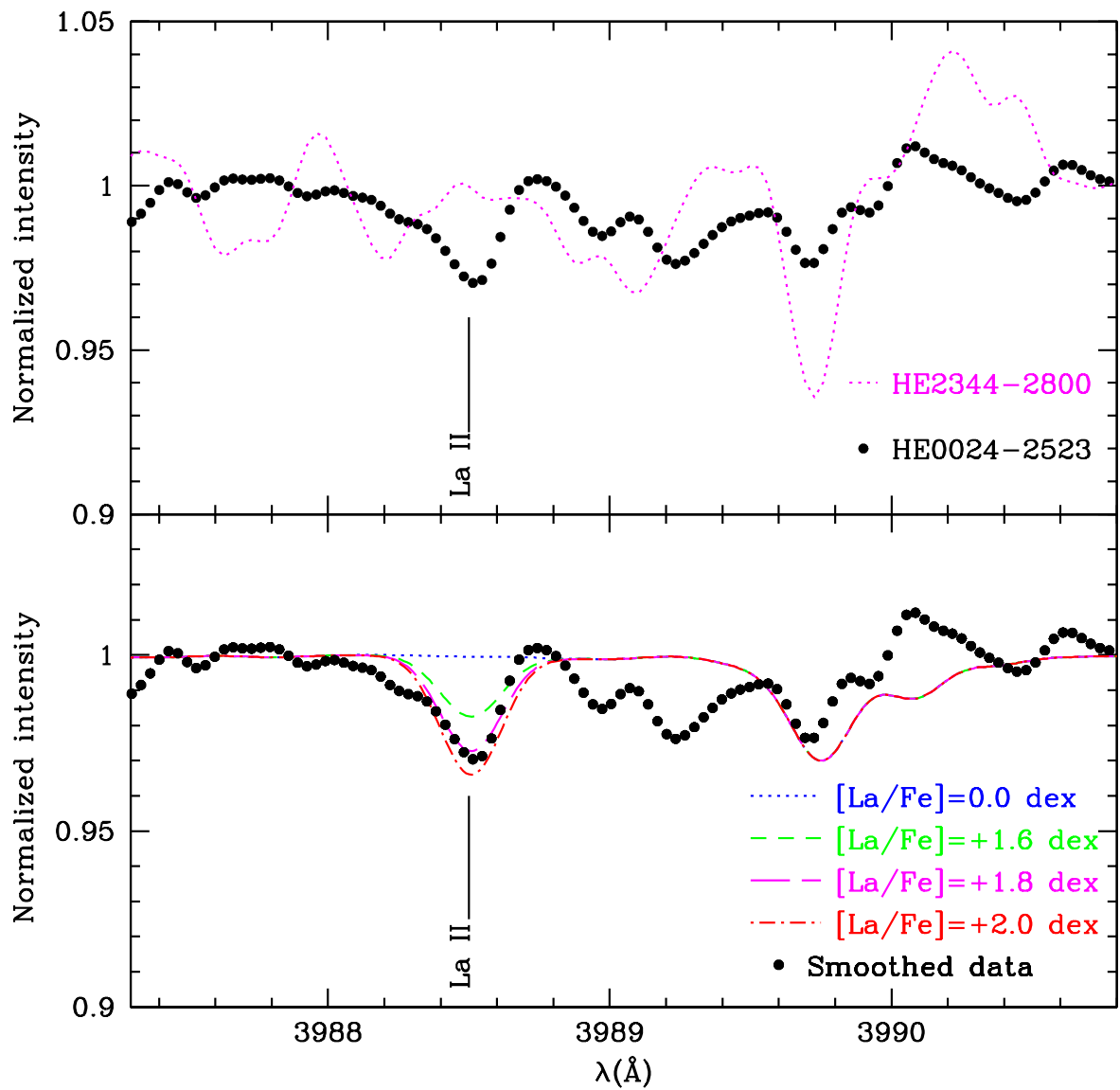}
\caption{Spectral region around the La II line at 3988.5\,\AA. The lower panel
shows the comparison of the smoothed spectrum of the program star with
synthetic ones computed for La abundances of  [La/Fe]=+1.6,  +1.8 and
+2.0\,dex. A synthetic spectrum with scaled solar La abundance is also
plotted. The signal-to-noise ratio in this spectral region is not very high.
In the upper panel we compare the spectrum of the program star with that of
the template star \CompStar, where the La feature is not detected.\label{la1}}
\end{figure}

\begin{figure}
\plotone{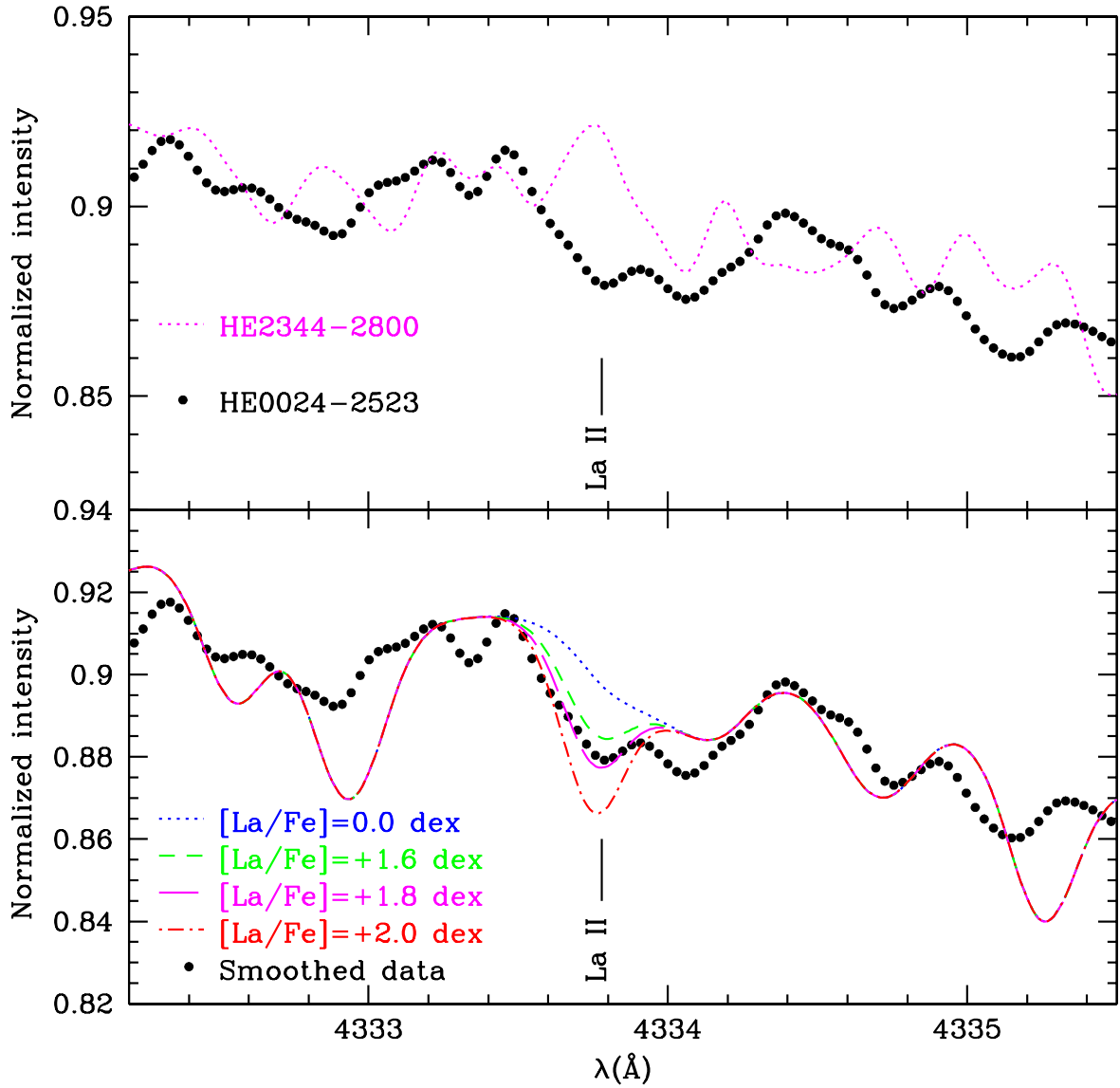}
\caption{Same as Figure \ref{la1} but for the spectral region around the La II
line at 4333.8\,\AA. The spectrum is quite noisy and the continuum is heavily
affected by the wings of the Balmer line H$_{\gamma}$ which is well
reproduced by our synthetic spectra. \label{la2}}
\end{figure}

\begin{figure}
\plotone{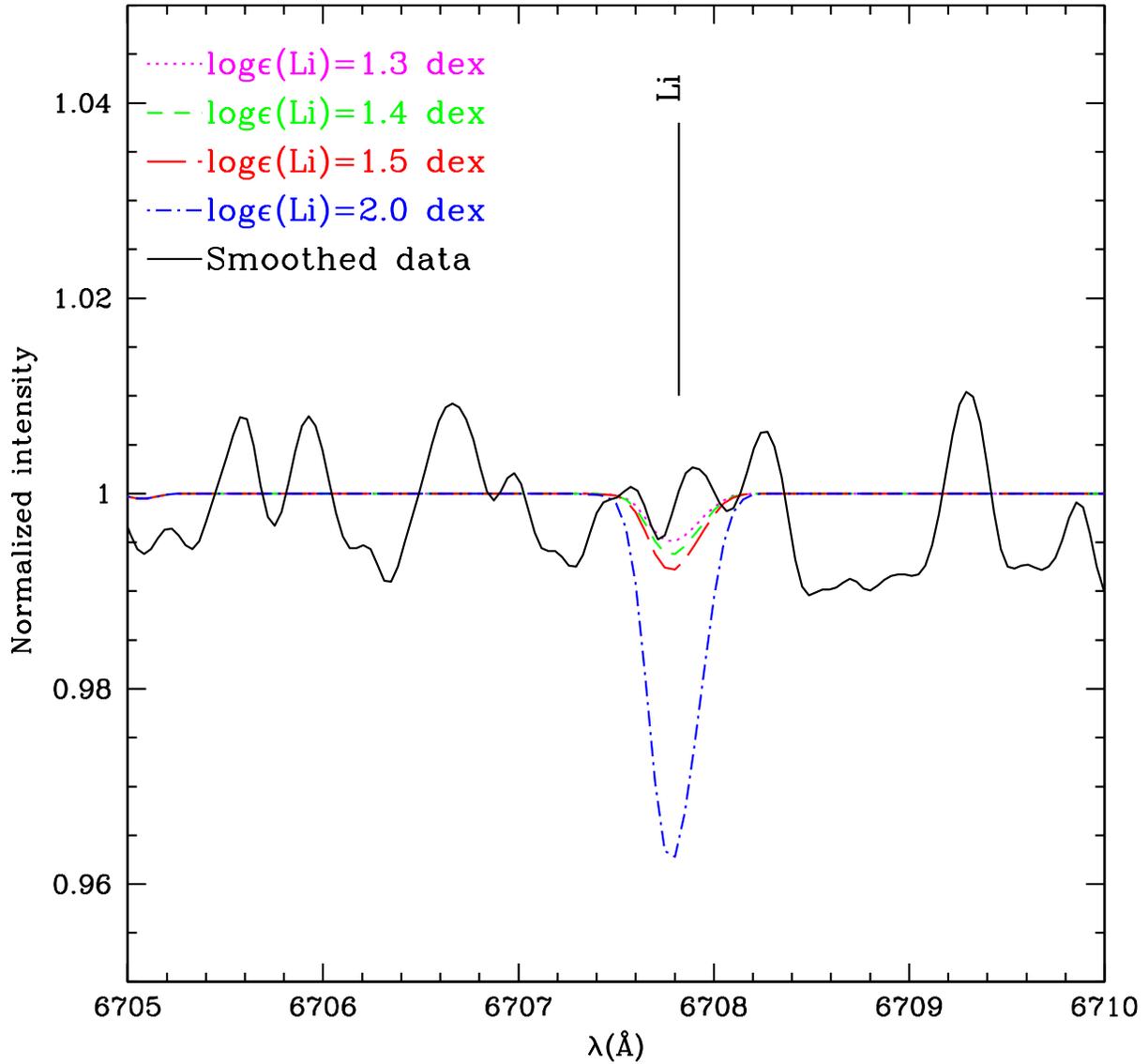}
\caption{Comparison between the observed data and the synthetic spectra for
the Li resonance line at 6707\,\AA. Notice the scale which is extremely blown
up. The synthetized spectra are for log$\epsilon$(Li)=+1.3, +1.4, +1.5 and
+2.0. The latter roughly represents the primordial Li abundance. The
difference between the expected intensity of the primordial Li feature and
that observed suggests that some Li depletion must have taken place.
\label{li}}
\end{figure}

\clearpage
\begin{figure}
\plotone{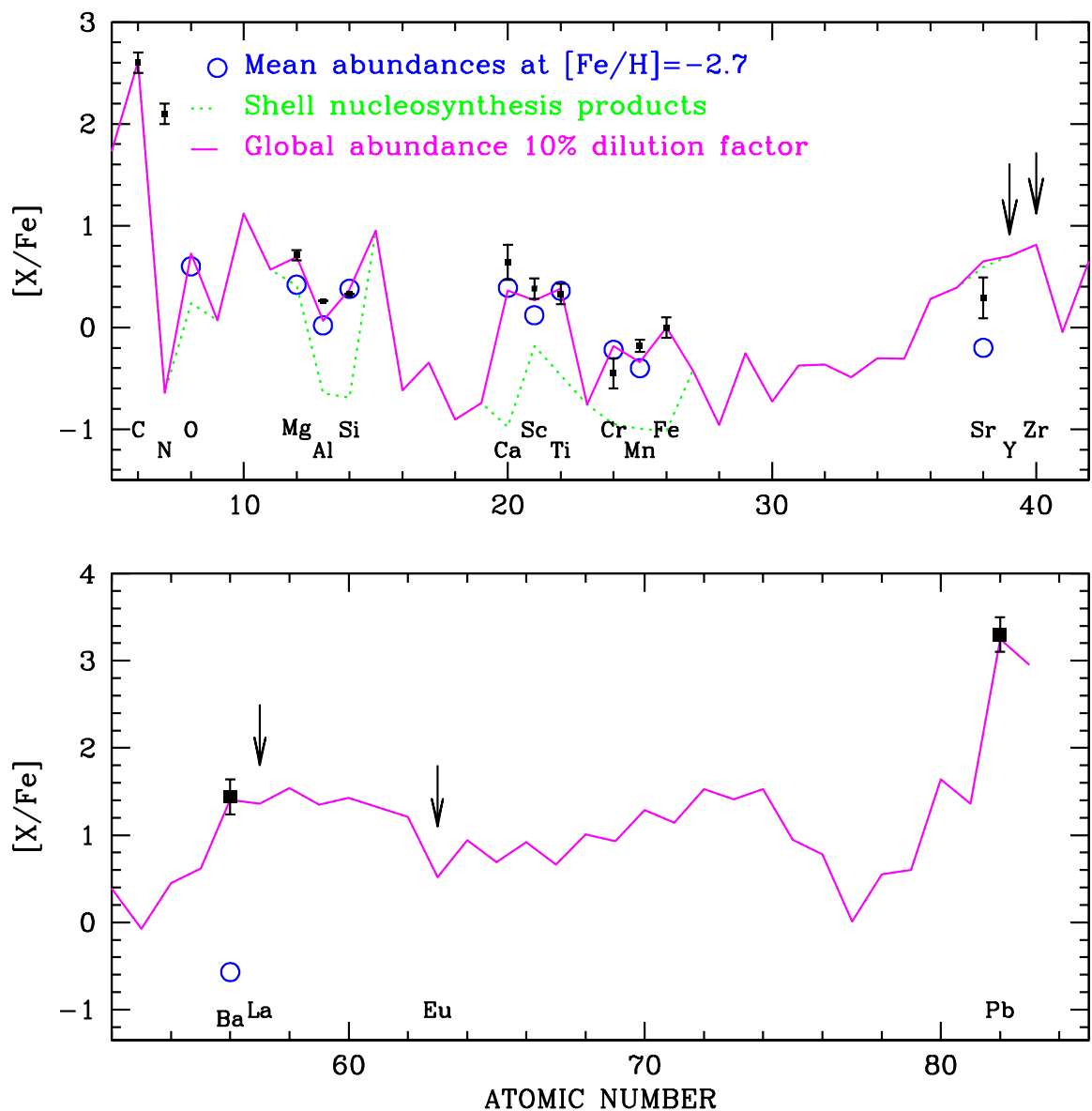}
\caption{ Comparison of our data with a nucleosynthesis model by Gallino
(priv. comm. 2002,  for details of the model see Busso  {\it et al.} 2001).
The model was calculated for a a metallicity of [Fe/H]=$-$2.6\,dex and  mass
M=1.5\,M$_{\odot}$. The dashed line represents the predicted shell
nucleosynthesis model abundance. The solid line is the resulting abundance for
the star's atmosphere after accretion of material from the companion assuming
a dilution factor of 10\%. The closed squares are our measured abundances,
while the arrows indicate the upper limits. The assumed original abundances of
the program star are indicated by the open circles (determined from the mean
observed abundances for stars at [Fe/H]=$-$2.7\,dex). The abundances have been
scaled to [Fe/H]=$-$2.7\,dex. \label{theory}}
\end{figure}

\clearpage

\begin{figure}
\plotone{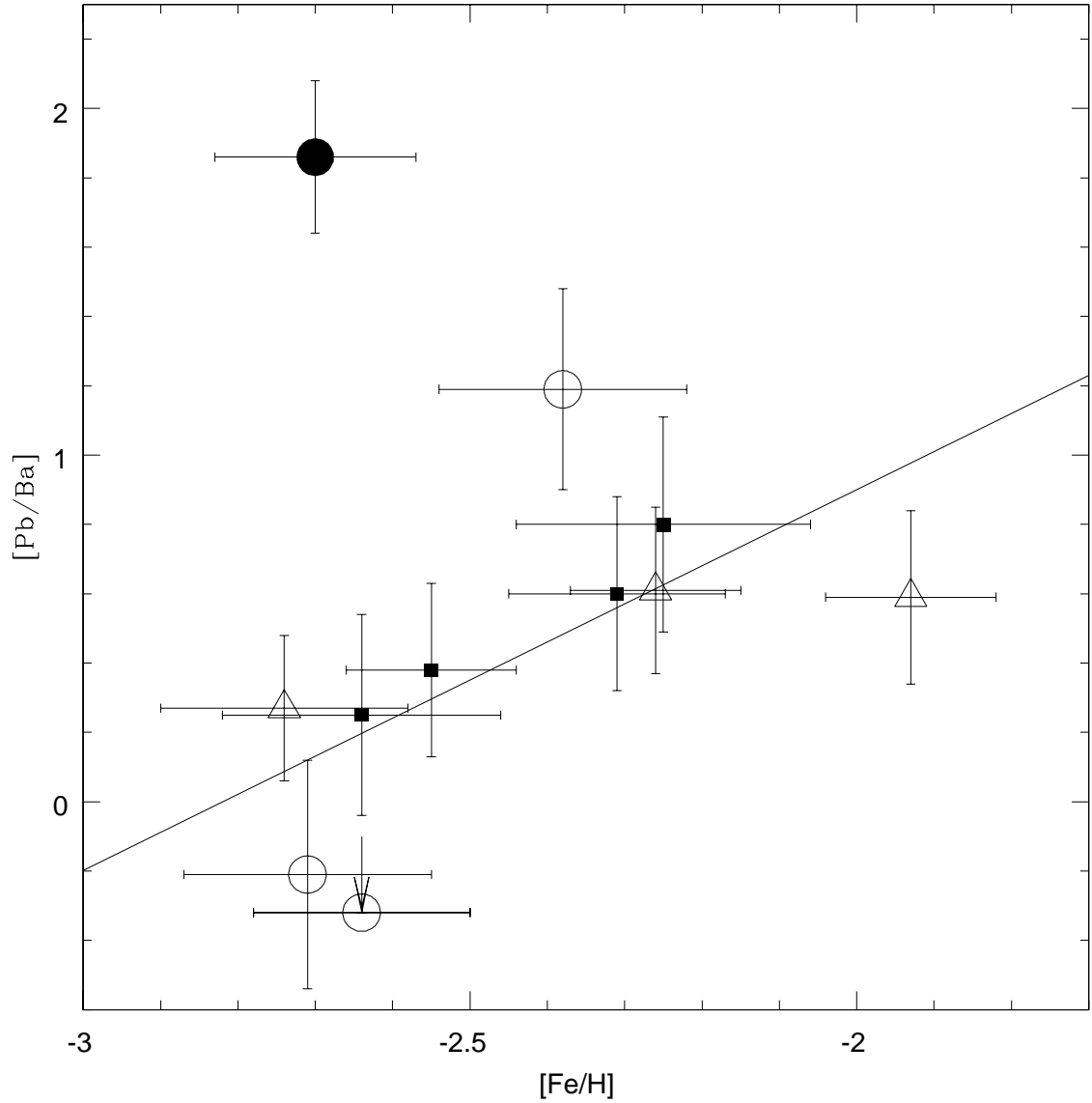}
\caption{[Pb/Ba] abundance as a function of [Fe/H] for \SarasStar\,and literature data (Aoki {\it et al.}
2002b,c and Aoki {\it et al.} 2000). \SarasStar\, is
represented with a closed circle. Stars with detected radial velocity variation are
represented by open circles, while those with no radial velocity variation over an interval
of several years (Preston \& Sneden 2001,
Norris, Ryan \& Beers 1997) are indicated with open triangles. Closed squares represent stars with a 
single observation. The addition of
\SarasStar\, to the sample seems to invalidate correlation between [Fe/H] and
[Pb/Ba] suggested by Aoki {\it et al.} (2002c). \label{comp}} 
\end{figure}

\clearpage
\begin{deluxetable} {lcccccc}
\tablewidth{0pt}
\tablecaption{Description of the observations. \label{obs}}
\tablehead{
\colhead{Date}&\colhead{Spec.$^{a}$}&
\colhead{Exp. time (s)}&\colhead{S/N$^{b}$}&
\colhead{MJD$^{c}$}&
\colhead{v$_{r}$(\kms)}&\colhead{$\sigma$(\kms)}}
\startdata
2000-09-24 & HIRES  &1200 &40&51811.386 &$-185.0$&0.4\\
2000-09-24 & HIRES  &1200 &40&51811.397 &$-183.6$&0.8\\
2000-09-24 & HIRES  &1200 &40&51811.412 &$-186.6$&0.7\\ 
2000-09-24 & HIRES  &1200 &40&51811.427 &$-187.4$&1.1\\
2000-09-24 & HIRES  &1200 &40&51811.442 &$-189.6$&0.3\\  
2001-09-27 & HIRES  &1200 &40&52179.398 &$-135.8$&1.1\\
2001-09-30 & HIRES  &1200 &40&52182.365 &$-125.6$&1.2\\
2001-08-27 & UVES   &3600 &80&52148.352 &$-127.0$&0.5\\
2001-10-23 & UVES   & 600 &30&52205.017 &$-218.2$&0.7\\
2001-10-23 & UVES   &1800 &40&52206.014 &$-136.2$&0.4\\
2001-10-25 & UVES   & 600 &20&52207.010 &$-160.8$&0.6\\
2001-10-26 & UVES   &1800 &35&52208.161 &$-228.6$&0.5\\
2001-10-28 & UVES   & 600 &20&52210.021 &$-133.9$&0.8\\
2001-10-28 & UVES   & 600 &20&52210.255 &$-148.2$&0.9\\
2001-10-25 & Palomar& 400 &30&52207.241 &$-186.2$&5.0\\
2001-10-26 & Palomar& 400 &30&52208.251 &$-227.8$&5.0\\
2001-10-27 & Palomar& 400 &30&52209.218 &$-145.6$&5.0\\
2001-10-28 & Palomar& 400 &30&52210.216 &$-140.1$&5.0\\ \hline
\enddata
\tablenotetext{a}{Spectrograph with which the observations were collected.}
\tablenotetext{b}{The signal-to-noise ratio is per resolution element.}
\tablenotetext{c}{Modified Julian Day  MJD=JD$-$2400000.5}
\end{deluxetable}

\clearpage

\begin{deluxetable} {lrrr}
\tablewidth{0pt}
\tablecaption{Our {\it best} orbital solutions. \label{orbsol}}
\tablehead{
\colhead{Orbital}&\colhead{Initial value for}&\colhead{Computed}&\colhead{Standard}\\ 
\colhead{element}&\colhead{last iteration}&\colhead{value}&\colhead{error}}
\startdata
&Circular Orbit & & \\
\hline \\
P (days)        &3.14  &3.41263&0.00008     \\
T$_{0}$\,(days)     &51811.0&52059.596&0.005    \\
G$^{a}$\,(\kms)       &0.0   &$-$178.3&0.4  \\
K$^{b}$\,(\kms)       &50.0  &51.9 &0.5  \\        
{\it e}         &0.0   &0.0 &\nodata \\ \hline
Projected semi-axis: & {\it a}$\sin i$\,(10$^{6}$\,km)             &2.438&0.023        \\
Mass-function: & f(m)\,(M$_{\odot}$) &0.0496&0.0014    \\  \hline
&&&\\
&Eccentric orbit&&\\
\hline
&&&\\
P (days)      &3.14 &3.41265&0.00016       \\
T$_{0}$\,(days)  &51811.0&52059.601&0.009       \\  
G$^{a}$\,(\kms)      &0.0  &$-$178.4&0.6            \\ 				       
K$^{b}$\,(\kms)      &50.0 &51.7&0.7              \\
{\it e}       &0.0  &0.010&0.016           \\
$\omega^{c}$            &0.0  &329   & 95             \\ \hline
Projected semi-axis: &{\it a}$\sin i$\,(10$^{6}$\,km) &2.43&0.03      \\
Mass-function: &f(m)\,(M$_{\odot}$)            &0.0490&0.0020      \\
\multicolumn{1}{l}{Epoch of ascending node:} & T\,(days)               &52062.7&0.9\\
\enddata
\tablenotetext{a}{Mean radial velocity of the system, {\it i.e.} radial velocity of the barycenter.}
\tablenotetext{b}{Semi-amplitude of the radial velocity curve.}
\tablenotetext{c}{Longitude of the ascending node.}
\end{deluxetable}

\clearpage

\begin{table}[h]
\caption{Comparison of the observed and predicted radial velocities. \label{vr}} 
\centering
\begin{tabular} {ccccr} \hline \hline
MJD&v$_{o}^{a}$  &Phase&v$_{c}^{b}$&(v$_{o}$-v$_{c}$)  \\
&(\kms)  &&(\kms)&(\kms)  \\
\hline
51811.39  & $ -185.0 $          &  ~~0.267     &$    -183.9   $ & $  ~-1.1$    \\
51811.40  & $ -183.6 $          &  ~~0.270     &$    -184.9   $ & $   +1.3$    \\
51811.41  & $ -186.6 $          &  ~~0.275     &$    -186.4   $ & $  ~-0.2$    \\
51811.43  & $ -187.4 $          &  ~~0.279     &$    -187.8   $ & $   +0.4$    \\
51811.44  & $ -189.6 $          &  ~~0.284     &$    -189.2   $ & $  ~-0.4$    \\
52148.35  & $ -127.0 $          &  ~~0.008     &$    -126.5   $ & $  ~-0.5$    \\
52179.40  & $ -135.8 $          &  ~~0.105     &$    -137.4   $ & $   +1.6$    \\
52182.37  & $ -125.6 $          &  ~~0.975     &$    -127.1   $ & $   +1.5$    \\
52205.02  & $ -218.2 $          &  ~~0.612     &$    -217.8   $ & $  ~-0.4$    \\
52206.01  & $ -136.2 $          &  ~~0.905     &$    -135.5   $ & $  ~-0.7$    \\
52207.01  & $ -160.8 $          &  ~~0.196     &$    -161.2   $ & $   +0.4$    \\
52208.16  & $ -228.6 $          &  ~~0.534     &$    -229.1   $ & $   +0.5$    \\
52210.02  & $ -133.9 $          &  ~~0.079     &$    -132.6   $ & $  ~-1.3$    \\
52210.26  & $ -148.2 $          &  ~~0.147     &$    -147.1   $ & $  ~-1.1$    \\
52207.24  & $ -186.2 $          &  ~~0.264     &$    -182.9   $ & $  ~-3.3$    \\
52208.25  & $ -227.8 $          &  ~~0.560     &$    -226.6   $ & $  ~-1.2$    \\
52209.22  & $ -145.6 $          &  ~~0.843     &$    -149.6   $ & $   +4.0$    \\
52210.22  & $ -140.1 $          &  ~~0.136     &$    -144.2   $ & $   +4.1$    \\ \hline \hline
\end{tabular}
\begin{list}{}{}
\item[]$^{a}$Observed radial velocity.\\
$^{b}$Computed radial velocity from our adopted solution.
\end{list}				          	               	                        
\end{table}

\clearpage
\begin{table}[h]
\caption{Mass of the {\it secondary} as a function of the mass of the {\it primary}.
\label{mass}}
\centering
\begin{tabular} {llllll} \hline \hline
\multicolumn{2}{c}{$\sin i=0.59$}&\multicolumn{2}{c}{$\sin i_{min}=0.40$}&\multicolumn{2}{c}{$\sin i_{max}=0.78$}\\
m$_{1}($M$_{\odot})$&m$_{2}($M$_{\odot})$&m$_{1}($M$_{\odot})$&m$_{2}($M$_{\odot})$&m$_{1}($M$_{\odot})$&m$_{2}($M$_{\odot})$\\ \hline 
      0.60  &    0.81  &    0.60  &    1.62  &    0.60   &   0.53\\
      0.65  &    0.83  &    0.65  &    1.66  &    0.65   &   0.55\\
      0.70  &    0.85  &    0.70  &    1.70  &    0.70   &   0.56\\
      0.75  &    0.87  &    0.75  &    1.74  &    0.75   &   0.58\\
      0.80  &    0.90  &    0.80  &    1.77  &    0.80   &   0.59\\
      0.85  &    0.92  &    0.85  &    1.81  &    0.85   &   0.60\\
      0.90  &    0.94  &    0.90  &    1.84  &    0.90   &   0.62\\
      0.95  &    0.95  &    0.95  &    1.88  &    0.95   &   0.63\\
      1.00  &    0.97  &    1.00  &    1.91  &    1.00   &   0.64\\
      1.05  &    0.99  &    1.05  &    1.94  &    1.05   &   0.65\\
      1.10  &    1.01  &    1.10  &    1.97  &    1.10   &   0.67\\
      1.15  &    1.03  &    1.15  &    2.00  &    1.15   &   0.68\\ \hline \hline
\end{tabular}
\end{table} 

\clearpage

\begin{deluxetable} {lllrrrl|lllrrrl}
\tablewidth{0pt}
\tablecaption{Equivalent widths for \SarasStar \label{ewt}}
\tabletypesize{\scriptsize}
\tablehead{
\colhead{Ion}& \colhead{$\lambda$} &
\colhead{$\chi$}& \colhead{$\log (gf)$} &
\colhead{Source$^{a}$}&\colhead{EW}& \colhead{weight$^{b}$}&
\colhead{Ion}& \colhead{$\lambda$} &
\colhead{$\chi$}& \colhead{$\log (gf)$} &
\colhead{Source}&\colhead{EW}&\colhead{weight$^{b}$}\\
\colhead{}& \colhead{(\AA)} &
\colhead{(eV)}& \colhead{} &
\colhead{}&\colhead{(m\AA)}& \colhead{}&
\colhead{(\AA)}& \colhead{(eV)} &
\colhead{}& \colhead{} &
\colhead{}&\colhead{(m\AA)}&\colhead{}}
\startdata
O I  &7771.95&  9.11& $  0.33 $&1 &   8.7&1&Fe I &3906.49&  0.11& $ -2.24 $&13&  27.5&1 \\
Na I &5889.95&  0.00& $  0.10 $&2 &  71.2&1&Fe I &3920.27&  0.12& $ -1.75 $&13&  55.5&1 \\
Na I &5895.92&  0.00& $ -0.20 $&2 &  45.6&1&Fe I &3922.92&  0.05& $ -1.65 $&13&  56.7&1 \\
Mg I &4057.52&  4.34& $ -1.20 $&3 &  13.3&1&Fe I &3930.31&  0.09& $ -1.59 $&22&  57.5&1 \\
Mg I &4167.28&  4.34& $ -1.00 $&3 &  10.8&0&Fe I &4005.24&  1.56& $ -0.61 $&15&  45.0&1 \\
Mg I &4703.00&  4.34& $ -0.67 $&3 &  19.9&1&Fe I &4045.81&  1.49& $  0.28 $&15&  76.5&1 \\
Mg I &5172.70&  2.71& $ -0.38 $&3 & 121.4&1&Fe I &4063.59&  1.56& $  0.08 $&19&  73.0&1 \\
Mg I &5183.62&  2.72& $ -0.16 $&3 & 142.7&1&Fe I &4071.74&  1.61& $ -0.02 $&15&  61.5&1 \\
Al I &3961.52&  0.00& $ -0.34 $&5 &  49.3&1&Fe I &4118.56&  3.57& $  0.14 $&19&  15.7&0 \\
Si I &3905.53&  1.91& $ -1.04 $&3a&  83.8&1&Fe I &4132.06&  1.61& $ -0.82 $&19&  42.1&1 \\
Ca I &4226.74&  0.00& $  0.24 $&4 & 103.7&1&Fe I &4143.87&  1.56& $ -0.62 $&19&  43.6&1 \\
Ca I &4425.44&  1.88& $ -0.36 $&4 &  13.6&1&Fe I &4187.05&  2.45& $ -0.55 $&16&  12.0&1 \\
Ca I &4435.69&  1.89& $ -0.52 $&4 &   8.6&1&Fe I &4187.81&  2.43& $ -0.55 $&16&  16.5&1 \\
Ca I &4454.79&  1.90& $  0.26 $&5 &  25.6&1&Fe I &4198.33&  2.40& $ -0.72 $&16&  17.8&0 \\
Sc II&4246.82&  0.32& $  0.32 $&22&  32.2&1&Fe I &4199.10&  3.05& $  0.16 $&19&  20.4&1 \\
Sc II&4314.08&  0.62& $ -0.10 $&22&  15.5&1&Fe I &4202.04&  1.49& $ -0.71 $&15&  37.7&1 \\
Sc II&4320.73&  0.60& $ -0.26 $&22&  12.3&1&Fe I &4227.44&  3.33& $  0.27 $&19&  11.2&1 \\
Ti I &3958.22&  0.05& $ -0.16 $&6 &  10.7&1&Fe I &4235.95&  2.43& $ -0.34 $&16&  27.4&1 \\
Ti I &3998.64&  0.05& $ -0.05 $&6 &  11.3&1&Fe I &4250.13&  2.47& $ -0.41 $&16&  12.9&1 \\
Ti I &4533.25&  0.85& $  0.48 $&7 &   6.4&1&Fe I &4250.80&  1.56& $ -0.38 $&19&  30.9&0 \\
Ti II&3900.54&  1.13& $ -0.45 $&22&  39.8&1&Fe I &4260.49&  2.40& $  0.14 $&18&  37.5&1 \\
Ti II&4012.39&  0.57& $ -1.61 $&22&  13.8&1&Fe I &4271.16&  2.45& $ -0.35 $&16&  24.2&1 \\
Ti II&4028.35&  1.89& $ -0.87 $&9 &   9.2&0&Fe I &4271.77&  1.49& $ -0.16 $&15&  59.4&1 \\
Ti II&4301.93&  1.16& $ -1.20 $&8 &  11.1&1&Fe I &4282.41&  2.18& $ -0.78 $&19&  10.3&1 \\
Ti II&4395.03&  1.08& $ -0.51 $&8 &  35.4&1&Fe I &4294.14&  1.49& $ -0.97 $&17&  36.6&1 \\
Ti II&4399.77&  1.24& $ -1.29 $&9 &  12.1&0&Fe I &4299.25&  2.43& $ -0.35 $&18&  23.5&1 \\
Ti II&4417.72&  1.16& $ -1.16 $&9 &  10.1&1&Fe I &4307.91&  1.56& $ -0.07 $&19&  65.9&1 \\
Ti II&4443.81&  1.08& $ -0.70 $&8 &  26.4&1&Fe I &4325.77&  1.61& $  0.01 $&19&  64.6&1 \\
Ti II&4468.51&  1.13& $ -0.60 $&8 &  31.4&1&Fe I &4383.56&  1.49& $  0.20 $&15&  73.0&1 \\
Ti II&4501.28&  1.12& $ -0.76 $&8 &  25.3&1&Fe I &4404.76&  1.56& $ -0.14 $&15&  61.7&1 \\
Ti II&4533.97&  1.24& $ -0.64 $&9 &  27.6&1&Fe I &4415.13&  1.61& $ -0.62 $&15&  40.3&1 \\
Ti II&4549.64&  1.58& $ -0.47 $&9 &  31.1&0&Fe I &4920.51&  2.83& $  0.15 $&19&  23.7&1 \\
Ti II&4563.77&  1.22& $ -0.82 $&9 &  20.0&1&Fe I &4957.61&  2.81& $  0.23 $&18&  40.4&1 \\
Ti II&4571.98&  1.57& $ -0.34 $&9 &  26.3&1&Fe I &5269.55&  0.86& $ -1.32 $&14&  52.8&0 \\
Ti II&4589.95&  1.24& $ -1.65 $&9 &   5.2&1&Fe II&4233.17&  2.57& $ -2.00 $&22&  17.4&1 \\
Cr I &4254.33&  0.00& $ -0.11 $&10&  13.3&1&Fe II&4583.84&  2.81& $ -2.02 $&22&  17.2&1 \\
Cr I &4274.79&  0.00& $ -0.23 $&10&  18.4&1&Fe II&4923.93&  2.88& $ -1.32 $&22&  18.3&1 \\
Mn I &4030.75&  0.00& $ -0.47 $&12&  23.4&1&Fe II&5018.45&  2.89& $ -1.22 $&22&  39.3&1 \\
Mn I &4033.06&  0.00& $ -0.62 $&12&  14.7&1&Sr II&4077.71&  0.00& $  0.15 $& 5&  69.7&1 \\
Fe I &3865.52&  1.01& $ -0.98 $&14&  48.5&1&Sr II&4215.52&  0.00& $ -0.17 $& 5&  53.1&1 \\
Fe I &3886.29&  0.05& $ -1.08 $&13&  70.9&1&Y  II&4883.69&  1.08& $  0.07 $&17& $<$5.9&1\\
Fe I &3887.06&  0.91& $ -1.14 $&14&  42.8&1&Zr II&4149.20&  0.80& $ -0.03 $&14& $<$9.0&1\\
Fe I &3895.67&  0.11& $ -1.67 $&13&  51.2&1&Ba II&4554.04&  0.00& $  0.17 $&21& 106.1&1\\ 
Fe I &3899.72&  0.09& $ -1.53 $&13&  46.8&1&Ba II&6141.75&  0.70& $  0.00 $&21&  56.5&1\\ 
Fe I &3902.96&  1.56& $ -0.47 $&15&  49.7&1&Ba II&6496.91&  0.60& $ -0.38 $&21&  34.2&1\\ 
\enddata		     
\tablenotetext{a} {The sources of $\log (gf)$ are as follows:
(1) Lambert \cite{lam}; (2) Biemont {\it et al.} \cite{bie};
(3) VALD database (as of August 2001); (3a) O'Brian,  T .~R. \& Lawler,  J.~E., \cite{obr};
(4) Smith \& Raggett \cite{smi}; (5) Wiese \& Martin \cite{wie};
(6) Blackwell {\it et al.} \cite{bla}; (7) Blackwell {\it et al.} \cite{blab};
(8) Bizzarri {\it et al.} \cite{biz}; (9) Magain \cite{mag}; 
(10) Blackwell {\it et al.} \cite{blac}; (11) Booth {\it et al.} \cite{booth}; 
(12) Blackwell {\it et al.} \cite{blad}; (13) Blackwell  {\it et al.} \cite{blae}; 
(14) Hannaford {\it et al.} \cite{han}; (15) Blackwell {\it et al.} \cite{blag}; 
(16) Blackwell  {\it et al.} \cite{blaf}; (17) Bi{\'e}mont {it et al.} \cite{biemo}; 
(18) Bard \& Kock \cite{bar}; (19) O'Brian {\it et al.} \cite{obri}; 
(21) Holweger \& M\"uller \cite{hol}; (22) Kurucz CD-ROM 23 \cite{kur2}}
\tablenotetext{b}{Weight attributed to the line, lines outside a 3 $\sigma$ interval were given zero weight}
\end{deluxetable}

\clearpage

\begin{deluxetable} {lrrrrr}
\tablewidth{0pt}
\tablecaption{Abundance summary for \SarasStar.
\label{abu}}
\tablehead{
\colhead{Element}     &  \colhead{n$^{a}$}
&  \colhead{Abundance} 
&\colhead{r.m.s.}&\colhead{Error}&\colhead{Method}\\
\colhead{ratio}     &  \colhead{}
&  \colhead{(dex)} 
&\colhead{(dex)}&\colhead{(dex)}&\colhead{}}
\startdata
{\rm [FeI/H]}           & 36& $-2.72$~~&0.12 & 0.12   &EWs\\
{\rm [FeII/H]}          &  3& $-2.67$~~&0.14 & 0.14   &EWs\\
{\rm log$\epsilon$(Li)} &   & $ 1.5$~~&\nodata& 0.1   &syn\\
{\rm [C/Fe]I}           &   & $ 2.6$~~&\nodata& 0.1   &syn\\
{\rm [N/Fe]I}           &   & $ 2.1$~~&\nodata& 0.1   &syn\\
{\rm [O/Fe]I}           &  1& $ 0.40^{b}$~~&\nodata&\nodata &EWs \\
{\rm [Na/Fe]I}          &  2& $-0.17^{b}$~~&0.13& 0.13   &EWs \\
{\rm [Mg/Fe]I}          &  4& $ 0.73$~~&0.11 & 0.06   &EWs \\
{\rm [Al/Fe]I}          &  1& $-0.35$~~&\nodata&\nodata &EWs \\
{\rm [Si/Fe]I}          &  1& $ 0.34$~~&\nodata&\nodata &EWs \\
{\rm [Ca/Fe]I}          &  4& $ 0.66$~~&0.13& 0.17   &EWs \\
{\rm [Sc/Fe]}II         &  3& $ 0.37$~~&0.14& 0.14   &EWs \\ 
{\rm [Ti/Fe]I}          &  3& $ 0.85$~~&0.13& 0.10   &EWs \\
{\rm [Ti/Fe]}II         & 12& $ 0.15$~~&0.06 & 0.05   &EWs \\
{\rm [Cr/Fe]I}          &  2& $-0.41$~~&0.21& 0.15   &EWs \\
{\rm [Mn/Fe]I}          &  2& $-0.26$~~&0.10& 0.06   &EWs \\
{\rm [Sr/Fe]}II         &  2& $ 0.34$~~&0.24& 0.24   &EWs \\
{\rm [Y/Fe]}II          &  1& $<0.91$~~&\nodata&\nodata&EWs \\
{\rm [Zr/Fe]}II         &  1& $<1.22$~~&\nodata&\nodata&EWs \\
{\rm [Ba/Fe]}II         &  3& $ 1.46$~~&0.20& 0.20   &EWs \\
{\rm [La/Fe]I}          &   & $ 1.8$~~&\nodata& 0.2   &syn\\
{\rm [Eu/Fe]I}          &   & $<1.1$~~&\nodata& 0.1   &syn\\
{\rm [Pb/Fe]I}          &   & $ 3.3$~~&\nodata& 0.1   &syn\\
{\rm $^{12}{\rm C}/^{13}{\rm C}$}& &  6   & \nodata&1      &syn\\
{\rm ~C/O}              &   &   100    &\nodata&\nodata &syn\\
{\rm ~N/O}              &   &    8     &\nodata &\nodata&syn\\
\enddata
\tablenotetext{a}{Number of lines over which the abundance was averaged}
\tablenotetext{b}{Includes non-LTE corrections according 
to the prescriptions of Gratton {\it et al.} \cite{gra}}
\end{deluxetable}

\clearpage

\begin{deluxetable} {lrrrrrrrrr}
\tablewidth{0pt}
\tablecaption{[Pb/Ba] ratios and identified periods for extremely metal poor, {\it s}-process
 enriched stars. \label{chs}}
\centering
\tablehead{\colhead{Star ID}&\colhead{[Fe/H]}&\colhead{$\sigma$}&\colhead{[Pb/Fe]}&\colhead{$\sigma$}
&\colhead{[Pb/Ba]}&\colhead{$\sigma$}&
\colhead{$^{12}$C/$^{13}$C}&\colhead{v$_{rad}^{a}$}&\colhead{Period}\\
\colhead{}&\colhead{(dex)}&\colhead{(dex)}&\colhead{(dex)}&\colhead{(dex)}
&\colhead{(dex)}&\colhead{(dex)}&
\colhead{}&\colhead{}&\colhead{(days)}}
\startdata
CS\,29526-110$^{1}$ &	$-$2.38&  0.16&	  3.30&  0.24  &   1.19 &  0.29&	  &	  yes &\\
CS\,22898-027$^{1}$ &	$-$2.26&  0.11&	  2.84&  0.19  &   0.61  & 0.24	&    $>$20$^{5}$:&        no &  \\
CS\,31062-012$^{1}$ &	$-$2.55&  0.11&   2.40&  0.19  &   0.38  & 0.25	&	  &           & \\ 
CS\,22880-074$^{1}$ &	$-$1.93&  0.11&	  1.90&  0.19  &   0.59  & 0.25	&    $>$40$^{5}$&        no & \\
CS\,31062-050$^{1}$ &	$-$2.31&  0.14&	  2.90&  0.24  &   0.60  & 0.28	&	  &           & \\       
CS\,22942-019$^{1}$ &	$-$2.64&  0.14&$<$1.60& \nodata&$<-$0.32 &\nodata&     30:$^{5}$&	   yes&2800$^{5}$ \\
HD\,196944 $^{1}$&	$-$2.25&  0.19&	  1.90&  0.24  &   0.80  & 0.31	&	  &           & \\ 
CS\,30301-015$^{1}$ &	$-$2.64&  0.18&	  1.70&  0.24  &   0.25  & 0.29	&	  &           & \\ 
HE\,0024-2523$^{2}$ & $-$2.72&  0.12&   3.30&  0.10  &    1.86 &  0.22&    	6 &       yes &3.14  \\ 
LP\,625-44$^{3}$    &	$-$2.72&  0.16&   2.60&  0.22  &  $-$0.21&  0.33&    	  &      yes  & \\
LP\,706-7$^{4}$     & $-$2.74&  0.16&   2.28&  0.20  &    0.27 &  0.21&         &       no  &\\ \hline 
\enddata
\tablenotetext{a}{Indicates whether radial velocity variation was detected.}
\tablerefs{(1) Aoki {\it et al.} 2002c; (2) Present work, (3) Aoki {\it et al.} 2002b;
(4) Aoki {\it et al.} 2001; (5) Preston \& Sneden 2001.} 
\end{deluxetable}

\end{document}